\documentclass[prb,twocolumn,preprintnumbers,amsmath,amssymb,superscriptaddress,showpacs,longbibliography,groupedaddress]{revtex4-1}
\usepackage{graphicx}
\usepackage{latexsym}
\usepackage{amsmath}
\usepackage{graphics}
\usepackage{amssymb}
\usepackage{layout}
\usepackage{verbatim}
\usepackage{amsfonts,epsfig,dsfont}
\usepackage{hyperref}

\newcommand{\ket}[1]{\left|#1\right>}
\newcommand{\bra}[1]{\left< #1 \right|}
\newcommand{\beq}{\begin{equation}}
\newcommand{\eeq}{\end{equation}}
\newcommand{\bea}{\begin{eqnarray}}
\newcommand{\eea}{\end{eqnarray}}

\newcommand{\mean}[1]{\langle{#1}\rangle{}}

\newcommand{\Vff}{\hat{V}_{\text{ff}}}

\newcommand{\MA}{\mathcal{A}}

\newcommand{\AM}{A_{\text{max}}}
\newcommand{\so}{\sigma_{\text{h}}}

\usepackage{color}

\usepackage{ulem} 

\begin{document}
\title{Dynamics of entanglement of two electron spins  interacting with nuclear spin baths in quantum dots}
\author{Igor Bragar}
\affiliation{Institute of Physics, Polish Academy of Sciences, al.~Lotnik{\'o}w 32/46, PL 02-668 Warszawa, Poland}
\author{{\L}ukasz Cywi{\'n}ski}
\email{lcyw@ifpan.edu.pl}
\affiliation{Institute of Physics, Polish Academy of Sciences, al.~Lotnik{\'o}w 32/46, PL 02-668 Warszawa, Poland}

\date{\today}

\begin{abstract}
We study the dynamics of entanglement of two electron spins in two quantum dots, in which each electron is interacting with its nuclear spin environment. Focusing on the case of uncoupled dots, and starting from either Bell or Werner states of two qubits, we calculate the decay of entanglement due to the hyperfine interaction with the nuclei.
We mostly focus on the regime of magnetic fields in which the bath-induced electron spin flips play a role, for example their presence leads to the appearance of entanglement sudden death at finite time for two qubits initialized in a Bell state. For these fields the intrabath dipolar interactions and spatial inhomogeneity of hyperfine couplings are irrelevant on the time scale of coherence (and entanglement) decay, and most of the presented calculations are performed using the uniform-coupling approximation to the exact hyperfine Hamiltonian. We provide a comprehensive overview of entanglement decay in this regime, considering both free evolution of the qubits, and an echo protocol with simultaneous application of $\pi$ pulses to the two spins. All the currently  relevant for experiments bath states are considered: the thermal state, narrowed states (characterized by diminished uncertainty of one of the components of the Overhauser field) of two uncorrelated baths, and a correlated narrowed state with a well-defined value of 
the $z$ component of the Overhauser field interdot gradient.
While we mostly use concurrence to quantify the amount of entanglement in a mixed state of the two electron spins, we also show that their entanglement dynamics can be reconstructed from measurements of the  currently relevant for experiments entanglement witnesses, and the fidelity of quantum teleportation performed using a partially disentangled state as a resource.  
\end{abstract}



\maketitle

\section{Introduction}
Qubits based on localized electronic spins in semiconductors have been a subject of intense research during the last 15 years. It is now possible to initialize, coherently manipulate, and measure spin qubits based on gated  quantum dots (QDs),\cite{Hanson_RMP07} self-assembled QDs,\cite{Liu_AP10,DeGreve_RPP13} nitrogen-vacancy (NV) centers in diamond,\cite{Dobrovitski_ARCMP13} and electrons bound to phosphorous donors in silicon.\cite{Zwanenburg_RMP13} Recently, advances in controlling the coupling between two qubits 
\cite{Petta_Science05,Robledo_Science08,Neumann_NP10,VanWeperen_PRL11,Kim_NP11,Greilich_NP11,Nowack_Science11,Brunner_PRL11} have led to demonstration of creation and manipulation of entangled states of two electron spin-based qubits.\cite{Shulman_Science12,Dolde_NP13} 

Quantum states of a single qubit suffer decoherence due to qubit's interaction with its environment.\cite{Zurek_RMP03} In many cases the dominant source of decoherence of semiconductor-based spin qubits is their coupling with other spins localized in the material. We focus here on the case in which this bath consists of spins of nuclei\cite{Coish_pssb09,Cywinski_APPA11,Urbaszek_RMP13,Chekhovich_NM13} of the atoms forming the material in which the qubit is localized, and the electron spin is coupled to the nuclei by the contact hyperfine (hf) interaction. 
The dynamics of a spin interacting with the nuclear bath is a theoretical problem that is more challenging than many of the model problems encountered in existing literature on entanglement dynamics.\cite{Carvahlo_PRL04,Mintert_PR05,Aolita_RPP15} The qubit--bath coupling 
is not necessarily weak in experimentally most interesting parameter regimes, the bath dynamics (caused both by the bath self-Hamiltonian, and by the qubit--bath interactions) is most often much slower than the timescale of qubit's decoherence, and the bath evolution can be strongly influenced by the presence and the dynamics of the qubit. All these observations preclude the use of Born-Markov approximation (i.e.~the use of Lindblad equation with time-independent coefficients).
The decoherence of a single qubit due to such an environment has been a subject of theoretical works employing a wide repertoire of approaches.
\cite{Merkulov_PRB02,Khaetskii_PRB03,deSousa_PRB03,Coish_PRB04,Coish_PRB05,Yao_PRB06,Witzel_PRB05,Witzel_PRB06,Zhang_PRB06,Chen_PRB07,Liu_NJP07,Coish_PRB08,Deng_PRB08,Yang_CCE_PRB08,Cywinski_PRL09,Cywinski_PRB09,Coish_PRB10,Cywinski_PRB10,Neder_PRB11,Barnes_PRB11,Barnes_PRL12,Sinitsyn_PRL12,Hung_PRB13,Stanek_PRB13,Faribault_PRB13,Beaudoin_PRB13,Balian_PRB14,Hackmann_PRB14,Welander_arXiv14,Dietl_PRB15} In experiments, many predicted key  features of nuclear-induced decoherence, such as the fast decay of coherence of a freely evolving qubit
\cite{Petta_Science05,Johnson_Nature05,Dutt_PRL05,Koppens_PRL08,Press_NP10} or oscillatory behavior of spin-echo signal in a QD with a few distinct species of nuclei,\cite{Bluhm_NP10} were observed. 
The existing level of agreement between experiment and theory allows us to assume that in the case of unstrained III-V QDs (i.e.~when the quadrupolar splittings\cite{Sinitsyn_PRL12,Chekhovich_NN12} of nuclear spins due to spatially inhomogeneous strain are negligible)
the microscopic description of the system is given by the Hamiltonian of contact hf coupling between the qubit and the bath spins, with the dipolar intrabath coupling\cite{deSousa_PRB03,Yao_PRB06,Witzel_PRB06}  playing a role only at long timescales. 

A pure entangled state of two qubits cannot be written as a tensor product of two pure single-qubit states, it has to be a superposition of at least two such product states.\cite{Horodecki_RMP09} 
As such it is expected to be at least as fragile as superpositions of single-qubit states when coupling with the environment is considered. Initially maximally entangled pure state evolves into a mixed, partially entangled state,\cite{Mintert_PR05,Aolita_RPP15} and the degree to which the state exhibits non-classical features (e.g.~breaking of Bell inequalities) or to which it overperforms classically correlated states at some tasks (such as quantum teleportation) is diminished. We are interested here in the dynamics of this entanglement decay process for two spin qubits interacting with nuclear baths.

We consider two localized electron spins (denoted by $A$ and $B$) playing the role of qubits, each of which can be coherently controlled and read out. 
The main example here  is the system of two electrons, each localized in a separate QD. The spins can then be initialized, controlled, and read out by either electrical\cite{Hanson_RMP07,Nowack_Science11,Brunner_PRL11} or optical\cite{Liu_AP10,DeGreve_RPP13} means, and the exchange interaction between spins in adjacent QDs, which was proposed years ago as means of entangling them,\cite{Loss_PRA98} can be controlled by tuning the gate voltages. The entanglement between distant QDs can be established by starting with an entangled state involving two proximal qubits, and then applying SWAP gates between consecutive pairs of dots along a chain 
(note however that for long chains this method might be susceptible to accumulation of gate errors and to unwanted effects due to finite residual exchange couplings between neighboring dots).
There are also theoretical proposals of methods allowing for long-range qubit--qubit coupling, 
\cite{Burkard_PRB06,Trifunovic_PRX13,Leijnse_PRL13} 
and also for moving of electrons along the chain of sites,\cite{Hollenberg_PRB06,Cole_PRB08} which should allow local creation of entanglement by nearest-neighbor exchange interaction to be followed by spatial separation of electrons. Long-distance shuttling of electrons with surface acoustic waves was recently achieved,\cite{Hermelin_Nature11,McNeil_Nature11} showing another way to obtain large spatial separation between entangled qubits. Finally, distant non-interacting qubits can be entangled in an entanglement swapping\cite{Zukowski_PRL93} protocol, in which two-qubit measurements in the Bell basis are used.

We focus on the case of III-V quantum dots, for which the hf-induced decoherence is unavoidable (since there are no nuclear spin-free isotopes of Ga, As, and In), and the experiments are often done at moderate magnetic fields (say, $B\! < \! 0.5$ T in GaAs).
In this work we  give a fairly detailed picture of entanglement dynamics of two electrons located in two uncoupled (possibly distant) QDs of this kind. 
Since the ability of the nuclear bath to cause electron spin flips is suppressed at finite magnetic fields, in the large field limit the influence of the bath amounts to pure dephasing, and calculation of entanglement decay of Bell states of uncoupled qubits becomes then a simple task involving only multiplying the known formulas for decaying single-qubit coherences. 
We focus thus on a more interesting (and experimentally relevant for gated GaAs QDs) regime of moderate magnetic fields, at which the electron-nuclear flip-flops influence visibly the entanglement dynamics, and the entanglement decay timescale is such that the effects of intrabath dipolar interactions\cite{Yao_PRB06,Witzel_PRB05,Witzel_PRB06} and the spatial inhomogeneity of hf couplings can be neglected (see the comparison between hf-only calculations\cite{Cywinski_PRL09,Cywinski_PRB09} and single qubit Hahn echo experiments done in this field regime\cite{Bluhm_NP10}).
We consider experimentally relevant states of the nuclear environments (thermal or narrowed, uncorrelated and correlated), and calculate the dynamics of two-spin entanglement both in the case of free evolution, and for the case of the two-spin echo protocol (which was employed in two recent experiments\cite{Shulman_Science12,Dolde_NP13} in which entanglement of spin-based qubits was demonstrated). As the initial states of the two electron spins we take the family of Bell-diagonal states, including the pure Bell states and the Werner state. We discuss the decay entanglement in all these cases using concurrence\cite{Wooters_PRL98} as means of quantification of amount of mixed state entanglement, but after establishing the general picture of the nature of disentanglement in all these cases, we discuss the time-dependence of other measures of two-qubit entanglement. Specifically, we calculate the decay of expectation values of easy-to-implement (for certain spin qubit systems) entanglement witnesses,
\cite{Guhne_PRA02,Guhne_PR09} and in particular we briefly describe the performance of the quantum teleportation protocol
\cite{Bennett_PRL93,Horodecki_PRA99,Verstraete_PRL03,vanEnk_PRA07} in which a partially disentangled state is used as a resource.  

Let us discuss briefly the relation of this paper to existing theoretical works on entanglement dynamics of electron spins interacting with nuclear spin baths. 
Decay of two-qubit entanglement in a high-temperature nuclear bath at large $B$ fields was discussed in Ref.~\onlinecite{Bodoky_JPC09}.
In Refs.~\onlinecite{Erbe_PRB10,Erbe_PRB12} the focus was on coupled qubits (non-zero exchange interaction between the two electron spins, leading to finite splitting of singlet and unpolarized triplet states, $\Delta_{\text{ST}}$), and magnetic field $B$ was assumed to be zero. 
Here we consider uncoupled spins ($\Delta_{\text{ST}} \! = \! 0$) and focus on $B \! \neq \! 0$ (with most attention devoted to regime of fields larger than the typical Overhauser field generated by the coupling to the nuclei). 
In two recent works\cite{Mazurek_PRA14,Mazurek_EPL14} both entanglement and more general quantum correlations were considered for two uncoupled spin qubits freely evolving at rather smalls $B$ fields (including $B\! =\! 0$), and interacting with thermal nuclear baths. Here we consider other bath states (narrowed, correlated), and we also calculate the entanglement dynamics under application of the spin echo protocol. 

Often the two QDs are situated one next to another 
in a double quantum dot (DQD) configuration. It is then possible to turn on the exchange coupling 
between the spins,\cite{Petta_Science05} and in many experiments such a DQD system was operated not as a pair of single-spin qubits, but as a host for a single logical qubit existing in total $S^{z}\! = \! 0$ subspace of the four-dimensional Hilbert space of two spins. 
The decoherence of the DQD-based singlet-triplet ($S$-$T_{0}$) qubit could be viewed as \textit{disentanglement} of a state of two single-spin qubits, since the relevant singlet and triplet states correspond to $\ket{\Psi_{\pm}}$ Bell states of the two spins. 
Of course, the fact that in the experiments on $S$-$T_{0}$ qubits only states from subspace spanned by $\ket{\Psi_{\pm}}$ are controlled explains why this point of view has not been stressed in literature. However, the decay of a $\ket{\Psi_{-}}$ state (i.e.~the decay of overlap of the decohered state with $\ket{\Psi_{-}}$, called ``singlet return probability'' in many experimental works) in coupled\cite{Laird_PRL06} and uncoupled\cite{Petta_Science05,Bluhm_NP10} DQDs is well understood theoretically. For decay at large $\Delta_{\text{ST}}$ see Ref.~\onlinecite{Coish_PRB05}, while the echo sequence for uncoupled QDs, with $\pi$ pulse provided by the two-qubit exchange gate, is considered in Ref.~\onlinecite{Neder_PRB11}, in which theory of Refs.~\onlinecite{Cywinski_PRL09,Cywinski_PRB09} is re-derived from a distinct point of view and applied to the $S$-$T_{0}$ case. The dephasing of superpositions of $S$ and $T_{0}$ states due to interaction with the nuclear bath was also considered theoretically for 
coupled dots in Refs.~\onlinecite{Yang_PRB08,Hung_PRB13} (note that the superposition of $S$ and $T_{0}$ can correspond to an entangled or a separable state depending on the value of the relative phase, with the simplest example being $(\ket{S}+\ket{T_{0}})/\sqrt{2} \! = \! \ket{\uparrow\downarrow}$ state which is separable, while $(\ket{S}+i \ket{T_{0}})/\sqrt{2}$ is maximally entangled).
In recent experiments on dephasing of such a $S$-$T_{0}$ superpositions in DQDs with two\cite{Weiss_PRL12,Dial_PRL13} and more
\cite{Higginbotham_PRL14} electrons, the dominant role of charge noise (leading to fluctuations of $\Delta_{\text{ST}}$) was uncovered, showing that in currently investigated structures the nuclear baths are not limiting the coherence at finite $\Delta_{\text{ST}}$.  

In the paper, we perform calculations for completely uncoupled QDs, that is we assume that the exchange splitting $\Delta_{\text{ST}} \! = \! 0$ even if the two qubits are in the DQD configuration. Of course $\Delta_{\text{ST}}$ is then not equal to zero exactly, but gate control of interdot detuning and barrier height allows for $\Delta_{\text{ST}}$ to be tunable in a wide range 
(see e.g.~Refs.~\onlinecite{Dial_PRL13,Higginbotham_PRL14} with recent results) encompassing the values of $\Delta_{\text{ST}}$ much smaller than typical energy of electron spin interaction with the thermal nuclear bath. For non-thermal baths and when spin echo protocol is used a conservative assumption of $\Delta_{\text{ST}}$ much smaller than the inverse characteristic timescale of coherence/entanglement decay can be used to quantify what is the realistic meaning of ``uncoupled'' qubits in the context of their entanglement dynamics. Note that measurements\cite{Bluhm_NP10} of echo decay in a GaAs DQD on timescales as long as $\approx\! 30$ $\mu$s were succesfully interpreted
\cite{Bluhm_NP10,Neder_PRB11} while assuming uncoupled dots inthe far-detuned regime of DQD operation.

In order to make this paper easy to read for both the practitioners of spin qubit physics, and the researchers who have experience with entanglement dynamics, but have not worked on nuclear bath, 
we have organized the paper in the following way. Section~\ref{sec:intro_spin} contains an attempt at a self-contained introduction to the most relevant features of the problem of hf-induced decoherence. It can be safely skipped (or only briefly scanned) by readers familiar with the topic. We note that the parameter regime on which we focus in this paper is discussed in Sec.~\ref{sec:moderate}.
In Sec.~\ref{sec:evolution} we describe the general features of dynamics (both for free evolution and for two-spin echo case) of the reduced density matrix of two spin qubits interacting with the nuclear spin bath, and we describe the approximation schemes used in this work. Section~\ref{sec:quantification} contains a short discussion of the issue of quantification of entanglement and calculation of its appropriate measures in the case of two-qubit entanglement. The calculations of two-qubit entanglement quantified by concurrence are presented in the following sections: in Sec.~\ref{sec:thermal} we show results for the case of the two uncorrelated nuclear baths in thermal equilibrium, while in Sec.~\ref{sec:narrowed} we consider the case of narrowed nuclear baths (i.e.~nuclear environments with a diminished uncertaintity in the value of one of the components of the effective field that the nuclei exert on the electron spin). Both the case of uncorrelated baths (each narrowed separately), and the experimentally 
relevant for DQDs case of correlated environments (with a well-defined value of the difference of the effective fields generated by the two nuclear ensembles) are considered there. Two-spin echo results are presented in Sec.~\ref{sec:echo}. Finally, in Sec.~\ref{sec:witnesses}, having previously discussed the general features of time-dependence of the two-qubit density matrix, we consider other ways of detecting and quantifying entanglement: using entanglement witnesses and performing a quantum teleportation protocol, both of which require some prior knowledge about the available mixed entangled state.

\section{Electron spins interacting with nuclear baths in quantum dots} \label{sec:intro_spin}
\subsection{Hamiltonian of the electron spins and of the spin-bath interaction}
The Hamiltonian of the electron spin interacting via contact hf coupling with nuclei in the QD is given by
\beq
\hat{H} = \Omega\hat{S}^{z} + \sum_{k}A_{k}\hat{\mathbf{S}}\cdot\hat{\mathbf{J}}_{k} \,\, ,\label{eq:Hhf}
\eeq
where $\Omega\! =\! g\mu_{B}B$ is the electron spin Zeeman splitting, in which $g$ is the effective $g$-factor of an electron in a QD, $\mu_{B}$ is Bohr's magneton, $B$ is the external magnetic field,
and $A_{k}$ is the hf coupling to the $k$-th nucleus.
In QDs based on III-V materials multiple kinds of nuclei are present (nuclei of various isotopes of Ga, As, In etc.). Denoting the nuclear species with index $\alpha$, 
the hf coupling for nucleus of species $\alpha$ located at site $k$ is $A_{k} \! = \! \mathcal{A}_{\alpha}|F(\mathbf{r}_{k})|^{2}$, where $F(\mathbf{r}_{k})$ is the envelope function of the localized electron state, and $\mathcal{A}_{\alpha}$ is the energy of hf interaction characteristic for a nucleus of a given atom embedded in a crystal lattice. 

It is convenient to define an effective number $N$ of nuclei  appreciably coupled to the electron spin by 
\beq
\frac{1}{N} \doteq \sum_{u}|F(\mathbf{r}_{u})|^{4} \,\, 
\eeq
where $u$ labels the primitive unit cells (and the envelope $F(\mathbf{r})$ is assumed to be approximately constant inside each cell). 
As we will see in Sec.~\ref{sec:bath}, an important role in the physics of spin-bath coupling is played by the quantity
\beq
\sum_{k}A^{2}_{k} \approx \sum_{\alpha} n_{\alpha}\mathcal{A}^{2}_{\alpha}\sum_{u}|F(\mathbf{r}_{u})|^{4} = \sum_{\alpha} \frac{n_{\alpha}\mathcal{A}^{2}_{\alpha}}{N} \,\, ,
\eeq
where $n_{\alpha}$ is the average number of nuclei of type $\alpha$ in the primitive unit cell. 

We can write the spin-bath coupling term as
\beq
\sum_{k}A_{k}\hat{\mathbf{S}}\cdot\hat{\mathbf{J}}_{k} \equiv \sum_{i=x,y,z} \hat{h}^{i}\hat{S}^{i} \equiv \hat{h}^{z}\hat{S}^{z} + \Vff \,\, ,  \label{eq:transVff}
\eeq
where we have introduced the three components of the Overhauser field operator, $\hat{h}^{i} \doteq \sum_{k} A_{k}J^{i}_{k}$, and in the second expression we have separated the term related to the $\hat{h}^{z}$ longitudinal component from the transverse components appearing in the electron-nucleus flip-flop operator $\Vff \! \doteq \! \frac{1}{2}\sum_{k}A_{k}(\hat{S}^{+}\hat{J}^{-}_{k} + \hat{S}^{-}\hat{J}^{+}_{k})$. In the next Section we will discuss the physical motivation for such a separation.

\subsection{Intrinsic bath dynamics and the nuclear density matrix relevant for possible kinds of experiments involving spin qubits} \label{sec:bath}
The bath Hamiltonian is given by
\beq
\hat{H}_{\text{bath}} = \sum_{k}\omega_{k}\hat{J}^{z}_{k} + \hat{H}_{\text{dip}} \,\, ,
\eeq
with  $\omega_{k}$ being the Zeeman splitting of the $k$-th nuclear spin, equal to one of $\omega_{\alpha}$ values depending on the kind of nucleus present at site $k$ (we assume here that the magnetic field is uniform on the length scale of the size of the QD; for theoretical treatment of the case of nonuniform field see e.g.~Ref.~\onlinecite{Beaudoin_PRB13}).
The second term is the internuclear dipolar interaction, which for typically used magnetic fields can be approximated by its secular form:
\beq
\hat{H}_{\text{dip}} = \sum_{i\neq j} b_{ij} ( \hat{J}^{+}_{i}\hat{J}^{-}_{j} -2  \hat{J}^{z}_{i}\hat{J}^{z}_{j} ) \,\, ,\label{eq:Hdip}
\eeq
where $b_{kl}$ are dipolar couplings between the nuclei. Their exact form is irrelevant here: what matters is that they decay rather quickly with distance, and that they are very weak. In all the experiments on spin qubits $b_{kl} \! \ll \! k_{B}T$, and also $\omega_{\alpha} \! \ll \! k_{B}T$ unless very high magnetic fields ($B \! > \! 10$ T) are used. This means that the equilibrium density matrix of the nuclei is 
\beq
\hat{\rho}^{\text{eq}}_{J} = \frac{1}{Z} \mathds{1} 
\,\, , \label{eq:rhoI}
\eeq
where the statistical sum $Z$ for a bath consisting of $N$ nuclei of spin $J$ is given by $Z\! =\! (2J+1)^{N}$. 
In some cases it is possible to treat the Overhauser field classically. The distribution of classical field $\mathbf{h}$ corresponding to the above density matrix is given by\cite{Merkulov_PRB02}
\beq
P(\mathbf{h}) = \frac{1}{(2\pi)^{3/2}\so^{3}} \exp \left ( -\frac{h^{2}}{2\so^{2}} \right) \,\, , \label{eq:Ph}
\eeq
where 
\beq
\so^{2} = \sum_{k}A^{2}_{k}\mean{(\hat{J}^{z}_{k})^2}_{\text{eq}} = \frac{1}{3} \sum_{\alpha}  J_{\alpha}(J_{\alpha}+1) n_{\alpha} \frac{\mathcal{A}^{2}_{\alpha}}{N} \,\, ,  \label{eq:sigma}
\eeq
where $\mean{...}_{\text{eq}}$ denotes averaging with respect to the nuclear density matrix from Eq.~(\ref{eq:rhoI}), and $J_{\alpha}$ is the length of spin of $\alpha$ species. The standard deviation $\so$ gives us the value of the typical effective Overhauser field felt by the electron spin interacting with a high-temperature nuclear bath.

Another consequence of weakness of internuclear dipolar interaction is the slowness of intrinsic nuclear dynamics. 
In the presence of $B$ field larger than the magnetic resonance linewidth of nuclei\cite{Abragam} (caused by $\hat{J}^{z}_{i}\hat{J}^{z}_{j}$ interactions in Eq.~(\ref{eq:Hdip})), i.e.~for $B \! \gg \! 0.1$ mT, the dynamics of longitudinal component of the Overhauser field, $h^{z}$, is much slower than that of the transverse components $h^{\perp} \! = \! h^{x}$, $h^{y}$.
The randomization $h^{\perp}$ occurs then on timescale of $\tau_{\perp} \! \sim \! 100$ $\mu$s given by the inverse of this linewidth, while the dynamics of $h^{z}$ is governed by diffusion of nuclear polarization (due to nearest-neighbor flip-flops caused by $J^{+}_{k}J^{-}_{l}$ interactions in Eq.~(\ref{eq:Hdip})) out of the volume of the QD.\cite{Deng_PRB05} The characteristic timescale $\tau_{||} \! \sim \! L^{2}/D$, where $L$ is the size of the QD and $D$ is the nuclear diffusion constant. The experiments give $\tau_{||}\! \sim\! 10$~s in GaAs QDs,
\cite{Reilly_PRL08} in qualitative agreement with theory.\cite{Deng_PRB05} 

In an experiment, the cycle of initialization, evolution, and measurement of the states of the qubits is repeated many times in order to gather enough data to reconstruct the time-dependence the electron density matrix. Often the acquisition of these data takes much longer time than $\tau_{||}$, so that we can safely assume ergodicity of the nuclear dynamics.
The measured signal corresponds then to averaging of the electron spin evolution over nuclear ensemble described by $\hat \rho^{\text{eq}}$ from Eq.~(\ref{eq:rhoI}). However, when single-shot readout methods are used,\cite{Barthel_PRL09} the qubit dynamics (e.g.~spin precession) can be measured on timescale order of magnitude shorter than $\tau_{||}$, but still much longer than $\tau_{\perp}$. The measurement of spin precession corresponds then to measurement of the full spin splitting, $\Omega+h^{z}$. Such a nuclear state is called the \textit{narrowed state},\cite{Coish_PRB04,Klauser_PRB06,Stepanenko_PRL06,Giedke_PRA06} and various degrees of narrowing were experimentally obtained in QDs using various methods.\cite{Bluhm_PRL10,Greilich_Science06,Shulman_NC14}
In the theoretical description\cite{Coish_PRB04,Yao_PRB06,Coish_PRB08,Cywinski_PRB09,Coish_PRB10,Barnes_PRB11,Barnes_PRL12} of such narrowed state free induction decay (NFID) experiments one should use a nuclear density matrix corresponding to a fixed value of $h^{z}$ (or at least a narrowed spread of possible values). 

We use the basis of bath states which are products of eigenstates of $J^{z}$ of individual nuclear spins, $\ket{\chi} = \ket{J^{z}_{1}...J^{z}_{N}}$, where $\hat{J}^{z}_{k}\ket{\chi} \! = \! J^{z}_{k}\ket{\chi}$, and we define $\ket{\chi(h^{z})}$ as a state corresponding to a given eigenvalue $h^{z}$ of $\hat{h}^{z}$. On the timescale of interest the nuclear environment is treated as a finite one, so we can treat $h^{z}$ as a discrete variable, and the number of $\ket{\chi(h^{z})}$ states corresponding to it, $M(h^z)$, is finite, therefore we can label these states $\ket{\chi_i(h^z)}$ with index $i \in [1,M(h^z)]$. A partially narrowed state can be written as
\beq
\hat{\rho}_{J}(\bar{h}^{z},\sigma_{\text{n}}) = \frac{1}{Z}\sum_{h^{z}} p_{\bar{h}^{z},\sigma_{\text{n}}}(h^{z}) \sum_{i}\ket{\chi_i(h^{z})}\bra{\chi_i(h^{z})} \,\, ,  \label{eq:rho_narrowed}
\eeq 
where $p_{\bar{h}^{z},\sigma_{\text{n}}}(h^{z})$ is a function peaked at $\bar{h}^{z}$ having width $\sigma_{\text{n}} \! \ll \! \so$, and the normalization constant $Z \! =\! \sum_{h^{z}} M(h^{z}) p_{\bar{h}^{z},\sigma_{\text{n}}}(h^{z})$. In the limit of perfect narrowing we have $p_{\bar{h}^{z},\sigma_{\text{n}}}(h^{z}) \! =\! \delta_{\bar{h}^{z},h^{z}}$.
Note that the lack of the off-diagonal elements in the above density operator is a consequence of the realistic assumption that the measurements are averaged over a time scale much larger than $\tau_{\perp}$. 

In most cases there is no correlation between the states of the nuclear baths in the two QDs, and the total initial density matrix of the environment is
\beq
\hat{\rho}_{J} = \hat{\rho}^{A}_{J}(0)\otimes \hat{\rho}^{B}_{J}(0) \,\, , \label{eq:rhouncorr}
\eeq
where the density matrices corresponding to nuclei in dots A and B are given either by Eq.~(\ref{eq:rhoI}) or (\ref{eq:rho_narrowed}). 
When dealing with a DQD, it is also possible to create a narrowed distribution of Overhauser field difference between the dots.
\cite{Foletti_NP09,Bluhm_PRL10,Shulman_NC14} This corresponds to a correlated state of the two nuclear environments with a diminished uncertaintity of $\Delta h^z\! \doteq \! h^{z}_{A}-h^{z}_{B}$:
\begin{align}
\hat{\rho}_{AB}(\Delta h^{z},\sigma_{\text{n}}) & = \frac{1}{Z_{AB}} \sum_{h^{z}_{A},h^{z}_{B}} p_{AB} \big(h^{z}_{A}-h^{z}_{B},\sigma_{\text{n}} \big)  \nonumber\\
& \!\!\!\!\!\!\!\!\!\!\!\!\!\!\!\!\!\!\!\!\!\!\!\!\!\!\!\!\!\! \times \sum_{i,j}
\big( \! \ket{\chi_i(h^{z}_{A})}\bra{\chi_i(h^{z}_{A})} \! \big) \otimes \big( \! \ket{\chi_j(h^{z}_{B})}\bra{\chi_j(h^{z}_{B})} \! \big) \,\, , \label{eq:rhocorr}
\end{align}
where $p_{AB}$ is peaked at $\Delta h^{z}$ while its width is $\sigma_{\text{n}}\! \ll \! \so$, and $Z_{AB} \! = \! \sum_{h^{z}_{A},h^{z}_{B}}M_A(h^{z}_{A}) M_B(h^{z}_{B}) p_{AB} \big(h^{z}_{A}-h^{z}_{B} \big)$. It is useful to note that the above can be written as
\beq
\hat{\rho}_{AB}(\Delta h^{z},\sigma_{\text{n}}) = \sum_{h^{z}_{A},h^{z}_{B}} w(h^{z}_{A}; \Delta h^{z}) \hat{\rho}_{A}(h^{z}_{A})\otimes  \hat{\rho}_{B}(h^{z}_{B}) \,\, , \label{eq:rhocorr_separable}
\eeq
where $\hat{\rho}_{Q}(h^{z}_{Q})$ is the perfectly narrowed state for dot $Q$ and the weights 
\begin{align}
w(h^{z}_{A}; \Delta h^{z}) &  = p_{AB}\big(h^{z}_{A}-h^{z}_{B},\sigma_{\text{n}} \big) \nonumber \\
& \times  M_A(h^{z}_{A}) M_B(h^{z}_{B} \! \! = \! \! h^{z}_{A} \! + \! \Delta h^{z})/Z_{AB} \,\, .
\end{align}

\subsection{Basics of decoherence due to interaction with the nuclear bath} \label{sec:decoherence_basics}
\subsubsection{Quasistatic bath and inhomogeneous broadening} \label{sec:QSBA}
From the above discussion it is clear that the nuclear bath belongs to a class of environments having slow fluctuations, leading to strongly non-Markovian features of qubits' coherence and entanglement dynamics. In fact, a reasonable zeroth-order approach to the problem of a freely evolving electron spin coupled to the nuclear bath is to treat the bath as \textit{static} on the timescale of qubit dynamics.
\cite{Taylor_QIP06,Cucchietti_PRA05,Cywinski_APPA11} In this quasistatic bath approximation (QSBA) decoherence occurs due to the averaging over various states of the environment, and the changes of these states occur \textit{between} the repetitions of the cycle of qubit initialization--evolution--measurement. 

For $\Omega \! \gg \! \so$ there is a well-defined quantization axis $z$, and the qubits initialized in superpositions of $\hat{S}^{z}$ eigenstates experience essentially pure dephasing process due to averaging of their precessions over a distribution of $h^z$ fields given by Eq.~(\ref{eq:Ph}). This leads to an inhomogeneously broadened decay of transverse spin components:
\beq
\mean{\hat{S}^{x,y}(t)} \propto \exp \! \Big( \! -(t/{T_{2}^{*}})^2 \Big) \,\, 
\eeq
where $T_{2}^{*} \! = \! \sqrt{2}/\so$ is of the order of $10$ ns in gated GaAs QDs. This decay is so fast that the dynamics of the bath, either intrinsic (due to dipolar interactions), or extrinsic (caused by interaction with the electron) can be neglected, making the calculation self-consistent.
The same statement holds in the case of partially narrowed state obtained in Ref.~\onlinecite{Bluhm_PRL10}, where $T_{2}^{*} \! \approx \! 100$ ns was obtained. The most recent experiments,\cite{Shulman_NC14} in which presumably partial narrowing led to $T_{2}^{*}\! \approx \! 1$ $\mu$s, remain to be analyzed more carefully in this context.

Note that for $\Omega \! \ll \! \so$ the electron spin decay is incomplete within the QSBA,\cite{Merkulov_PRB02} and the description of $\mean{S^{x,y,z}(t)}$ decay towards zero requires a theoretical approach which takes into account the detailed shape of the electron's wavefunction.\cite{Zhang_PRB06,Chen_PRB07,Faribault_PRB13,Stanek_PRB13,Hackmann_PRB14} However, it is established\cite{Zhang_PRB06} that the very low-field QSBA solution correctly describes the exact dynamics of the system on timescale of $t \! \lesssim \! T_{2}^{*}$. 

\subsubsection{Looking beneath inhomogenous broadening -- very high magnetic fields} \label{sec:beneath}
There are two established methods of removing the inhomogenous broadening effect: the previously discussed creation of narrowed state of the nuclei and the spin echo experiment. In the latter case each qubit is rotated by $\pi$ about one of the in-plane ($x$ or $y$) axes at the midpoint of the evolution (at time $\tau$), and the coherence signal experiences that a revival (an echo) at the measurement time $2\tau$. The essence of the echo sequence is a complete removal of the influence of all the low-frequency ($\lesssim \! 1/(2 \tau)$) fluctuations of the $h^{z}$ field affecting the qubit's energy splitting.\cite{deSousa_TAP09,Cywinski_PRB08} 

Both of these methods remove the influence of fluctuations of $h^{z}$ which have characteristic timescale much longer than the time of single evolution of the qubit. The decoherence is then caused either by faster fluctuations of $h^{z}$, or by the influence of transverse Overhauser fields (i.e.~by the $\Vff$ term from Eq.~(\ref{eq:transVff})).
The latter effect is however diminishing with increasing $\Omega$: due to the fact that $\Omega$ is about $1000$ times larger than $\omega_{\alpha}$, for $\Omega \! \gg \! \so$ (when the probability that the $h^{z}$ is compensating the external field is vanishing) the $\Vff$ operator can lead only to virtual higher-order transitions between the states of the system. In the semiclassical picture the influence of the transverse field $|h_{\perp}| \! \sim \!  \so$ is strongly suppressed for $\Omega \! \gg \! \so$, and this fields leads then only to precession frequency renormalization by $\sim \! h^{2}_{\perp}/2\Omega$ and the tilt of quantization axis by angle $\sim \! h_{\perp}/\Omega$. At large enough $\Omega$ these effects become irrelevant, and the high-field coherence is limited by the intrinsic dynamics of $h^z$ caused by dipolar interactions.\cite{deSousa_PRB03,Witzel_PRB05,Yao_PRB06,Witzel_PRB06} The single-spin coherence decay can be described by taking into account the decohering influences of finite-
size groups (clusters) of nuclear spins\cite{Witzel_PRB05,Yao_PRB06,Witzel_PRB06,Yang_CCE_PRB08}. Such cluster expansion theories give
\beq
\mean{\hat{S}^{x,y}(t)} \propto \exp \! \Big( \! -(t/T_{2})^{\gamma} \Big) \,\, ,  \label{eq:Sxy_dip}
\eeq
with $\gamma \! = \! 4$ and $T_{2} \! \approx \! 10-50$ $\mu$s (depending on the shape of the wavefunction\cite{Witzel_PRB08}) for GaAs in the NFID and echo. 
This pure dephasing process sets the limit for the coherence time of an electron interacting with the nuclear bath.

\subsubsection{Looking beneath inhomogenous broadening -- low and moderate magnetic fields} \label{sec:moderate}
With decreasing $\Omega$  the role of the $\Vff$ term increases, and at some $\Omega$ (corresponding to  $B \sim 0.5$~T in GaAs QDs when spin echo decay is considered\cite{Bluhm_NP10}) the influence of $\hat{h}_{\perp}$  replaces the dipolar-induced fluctuations of $\hat{h}^z$ as the dominant source of decoherence. From multiple approaches
\cite{Coish_PRB04,Yao_PRB06,Coish_PRB08,Deng_PRB08,Cywinski_PRL09,Cywinski_PRB09,Coish_PRB10,Cywinski_PRB10,Neder_PRB11,Barnes_PRB11,Barnes_PRL12} to this problem of purely hf-induced decoherence we choose the ones that are expected to work at these moderate $B$ fields.
Specifically, we concentrate on field regime in which the decay of either NFID or echo occurs at times $t\! \ll \! 1/\AM$, where $\AM$ is the maximal value of the hf coupling $A_{k}$. In III-V QDs, for which there is a nuclear spin at every lattice site, and for which all the $\mathcal{A}_{\alpha}$ have similar values, we have $1/\AM \! \approx \! N/\mathcal{A} \! \approx \! 10$ $\mu$s for $N\! = \! 10^{6}$. For $t\! \ll \! 1/\AM$ the time-energy uncertainty should make the exact shape of $A_{k}$ distribution irrelevant, allowing us to  take all the $A_{k}$ couplings to be the same. 
If we can also restrict ourselves to regime of $t\! \ll \! \text{min}_{k,l}[1/(\omega_{k}-\omega_{l})]$ where $\omega_{k}$ and $\omega_{l}$ are Zeeman splittings of distinct nuclei (this condition is also easier to fulfill at low $B$), we can neglect the presence of mutliple nuclear species, and treat the nuclear bath as a large single spin $\mathbf{\hat{J}} \! = \! \sum_{k=1}^{N}\mathbf{\hat{J}}_{k}$.
Such a uniform coupling (UC) model can be easily solved exactly both in the case of free evolution, including NFID\cite{Barnes_PRB11}, and in the case of spin echo \cite{Cywinski_PRB10} for practically any $N$, and for any value of $\Omega$, including $\Omega\! =\! 0$. In the context of entanglement dynamics it is important that this method allows for calculation of all the components of $\hat{\rho}(t)$, since certain features of entanglement decay, for example the presence or absence of entanglement sudden death\cite{Zyczkowski_PRA01,Yu_PRL04} (ESD) when one of the Bell states is considered, can be captured only with a theoretical approach that describes the changes of qubit's populations. 

When dynamics on longer time scales is of interest, an approach that takes into account the presence of distinct nuclear species and/or the inhomogeneity of hf couplings becomes necessary. This is especially true for the case of spin echo, in which the crucial role of presence of nuclear species with distinct $\omega_{k}$ was predicted\cite{Cywinski_PRL09,Cywinski_PRB09} and verified experimentally.\cite{Bluhm_NP10} The theory used in these papers was based on employing an effective pure dephasing Hamiltonian derived in the second order with respect to $\Vff$. A brief outline of this method is given below in Sec.~\ref{sec:RDT}.

\section{Evolution of the reduced density matrix}  \label{sec:evolution}
We make two basic assumptions here. The first is that of  a lack of system--environment correlation at initial time, i.e.~we take the initial total density matrix to be 
$\hat{\rho}(0)\otimes\hat{\rho}_{J}(0)$, where $\hat{\rho}(0)$ is the initial density matrix of the qubits. The second is that there is no interaction between the two electron spins, and the two nuclear environments are well-separated, i.e.~even if the two dots are close to each other, the overlap of the wavefunctions of the two electrons is negligible, so that there is no exchange coupling, and each nuclear spin appreciably coupled to the electrons can be assigned to only one of the two baths. Formally, the total Hamiltonian of the system can be written as $\hat{H}_{A}\otimes \mathds{1}_{B} + \mathds{1}_{A} \otimes \hat{H}_{B}$, 
where $\hat{H}_{Q}$ is the Hamiltonian of the electron and the nuclei in dot $Q$.

\subsection{General formalism in the case of free evolution} \label{sec:evo_FID}
An important property of $\hat{H}_{Q}$ consisting both of the Hamiltonian of isotropic hf interaction from Eq.~(\ref{eq:Hhf}) and the intrabath dipolar interaction from Eq.~(\ref{eq:Hdip}) is the conservation law
\beq
[\hat{H}_{Q},\hat{S}_{Q}^{z}+\sum_{k\in Q}\hat{J}^{z}_{k} ] = 0 \,\, .   \label{eq:cons_law}
\eeq
We introduce operators $\Pi_{m,\sigma}\! \doteq \! \hat{P}_{\sigma\sigma}\hat{\Pi}_{m}$, where $\hat{P}_{\sigma\sigma'} \! \doteq \! \ket{\sigma}\bra{\sigma'}$ for $\sigma\! = \! \sigma'$ is projecting on subspace of given electron spin $S^{z}\!=\!\sigma/2$ (where $\sigma \! = \! \pm 1$), and $\hat{\Pi}_{m}$ is projecting on subspace of fixed total $z$ component of the nuclear spin, 
i.e.~the subspace of states $\ket{\chi(m)}$ fulfilling $\sum(k)\hat{J}^{z}_{k} \ket{\chi(m)} \! =\! m \ket{\chi_{m}}$. The the evolution operator $\hat{U}_{Q}(t) \! \doteq \exp(-i\hat{H}_{Q}t)$ (note that we put $\hbar \! =\! 1$) fulfills 
\beq
\Pi_{m,\sigma}\hat{U}_{Q}(t) \Pi_{m',\sigma'} = \delta_{m+\sigma/2,m'+\sigma'/2} \Pi_{m,\sigma}\hat{U}_{Q}(t) \Pi_{m',\sigma'} \,\, . \label{eq:PiU}
\eeq
This means that evolution starting from a state in the $\{\sigma,m\}$ subspace  leads to states located either in this subspace, or in the $\{\bar{\sigma},m+\sigma/2\}$ subspace (with $\bar{\sigma} \! \equiv \! -\sigma$). This property allows for an exact solution
\cite{Khaetskii_PRB03} in the case of fully polarized bath (and any distribution of $A_{k}$ couplings), but  it is also useful in the general case.

A general formula for the evolution of the two-qubit density matrix is
\beq
\rho_{\alpha\beta,\gamma\delta}(t) = \sum_{\alpha'\beta'\gamma'\delta'} R^{\alpha'\beta'\gamma'\delta'}_{\alpha\beta\gamma\delta} \! (t) \, \rho_{\alpha'\beta',\gamma'\delta'}(0) \,\, ,  \label{eq:evo_general}
\eeq
with
\beq
R^{\alpha'\beta'\gamma'\delta'}_{\alpha\beta\gamma\delta}(t) \doteq \text{Tr}_{J} \! \left( \bra{\alpha\beta} \hat{U}(t)\ket{\alpha'\beta'} \hat{\rho}_{J}(0)  \bra{\gamma'\delta'} \hat{U}^{\dagger}(t)\ket{\gamma\delta} \right ) \,\, ,     \label{eq:Rabcd}
\eeq
where $\hat{U}(t) \! \doteq \! \hat{U}_{A}(t)\otimes\hat{U}_{B}(t)$, and $\ket{\alpha\beta}\! \doteq \! \ket{\alpha}_{A}\otimes\ket{\beta}_{B}$ being the two-qubit state.

Let us start with a rather general density matrix of the environment $\hat{\rho}_{J}$: the one from Eq.~(\ref{eq:rhocorr}) but additionally with $p(h^{z}_{A}-h^{z}_{B})$ replaced by  $p(h^{z}_{A},h^{z}_{B})$, i.e.~we consider now a very broad class of possibly correlated environmental states. 
Using Eq.~(\ref{eq:PiU}) and the fact that $\Pi_{m_{Q}} \hat{\rho}_{J} \Pi_{n_{Q}} \! \propto \delta_{m_{Q}n_{Q}}$, we arrive at
\begin{align}
R^{\alpha'\beta'\gamma'\delta'}_{\alpha\beta\gamma\delta}(t) & \equiv \delta_{\alpha\alpha'}\delta_{\beta\beta'}\delta_{\gamma\gamma'}\delta_{\delta\delta'} R^{\alpha\beta\gamma\delta}_{\alpha\beta\gamma\delta}(t) \nonumber \\
& + \delta_{\alpha\alpha'} \delta_{\bar{\beta}\beta'} \delta_{\gamma\gamma'} \delta_{\beta\delta}\delta_{\bar{\beta}\delta'} R^{\alpha\bar{\beta}\gamma\bar{\beta}}_{\alpha\beta\gamma\beta}(t) \nonumber \\
& + \delta_{\bar{\alpha}\alpha'} \delta_{\beta\beta'} \delta_{\delta\delta'} \delta_{\alpha\gamma}\delta_{\bar{\alpha}\gamma'} R^{\bar{\alpha}\beta\bar{\alpha}\delta}_{\alpha\beta\alpha\delta}(t) \nonumber \\
& + \delta_{\alpha\gamma} \delta_{\bar{\alpha}\alpha'} \delta_{\bar{\alpha}\gamma'} \delta_{\beta\delta} \delta_{\bar{\beta}\beta'} \delta_{\bar{\beta}\delta'} R^{\bar{\alpha}\bar{\beta}\bar{\alpha}\bar{\beta}}_{\alpha\beta\alpha\beta}(t) \,\, . \label{eq:Rabcd4}
\end{align}
We can see how the constraints on the nuclear density matrix discussed in Sec.~\ref{sec:bath} and the conservation law (\ref{eq:cons_law}) reduce the number of $R$ evolution functions needed for description of two-qubit dynamics.

For uncorrelated baths from Eq.~(\ref{eq:rhouncorr}) we can express the $R$ functions defined above using the  functions describing the evolution of a \textit{single} qubit interacting with its bath.\cite{Bellomo_PRL07,Bellomo_PRA08} The evolution of qubit $Q$ is given by:
\begin{eqnarray}
\rho^{Q}_{\sigma\sigma'}(t) & = & \text{Tr}_{J} \! \left( \bra{\sigma}\hat{U}_{Q}(t) \hat{\rho}^{Q}(0)\otimes\hat{\rho}^{Q}_{J}(0) \hat{U}_{Q}^{\dagger}(t) \ket{\sigma'} \right ) \,\,  , \nonumber \\
& = & \sum_{\xi,\xi'} \text{Tr}_{J} \! \left( \bra{\sigma}\hat{U}_{Q}(t)\ket{\xi} \hat{\rho}^{Q}_{J}(0) \bra{\xi'}\hat{U}_{Q}^{\dagger}(t) \ket{\sigma'} \right ) {\rho}^{Q}_{\xi\xi'}(0) \,\, , \nonumber\\
& \equiv & \sum_{\xi,\xi'} K^{Q}_{\sigma\xi,\xi'\sigma'}(t) \, {\rho}^{Q}_{\xi\xi'}(0) \,\, , \label{eq:Kdef}
\end{eqnarray}
where $K^{Q}_{\sigma\xi,\xi'\sigma'}(t)$ encapsulates the influence of the environment on the evolution of the reduced density matrix of qubit $Q$. The above representation\cite{Bellomo_PRL07,Bellomo_PRA08,Mazurek_PRA14} is of course closely related to the operator-sum (Kraus) representation of evolution of $\hat{\rho}^{Q}$, but it is much more convenient to use when the number of relevant Kraus operators is large (as it is the case here). The $K$ functions have the following structure:
\beq
K_{\sigma \xi, \xi' \sigma'} \equiv \delta_{\sigma \xi} \delta_{\sigma' \xi'} K^{\sigma \sigma'}_{a} + \delta_{\sigma \sigma'} \delta_{\sigma \bar{\xi}} \delta_{\sigma \bar{\xi}'} K^{\sigma}_{b} \,\, ,  \label{eq:Kab}
\eeq
with
\begin{eqnarray}
K^{\sigma\sigma'}_{a} & \doteq & \sum_{m} \text{Tr} \! \left( \hat{P}_{\sigma'\sigma}\Pi_{m}\hat{U}\Pi_{m}\hat{P}_{\sigma\sigma'}\rho_{J}\Pi_{m}\hat{U}^{\dagger} \right ) \,\, , \label{eq:Ka} \\
K^{\sigma}_{b} & \doteq & \sum_{m} \text{Tr} \! \left( \hat{P}_{\sigma\sigma}\Pi_{m}\hat{U}\Pi_{m+\sigma/2}\hat{P}_{\bar{\sigma}\bar{\sigma}}\rho_{J}\Pi_{m+\sigma/2}\hat{U}^{\dagger} \right ) \,\, ,  \label{eq:Kb} 
\end{eqnarray}
where the time dependence of $K_{a/b}(t)$ and $\hat{U}(t)$ has been suppressed. 
The evolution of $\hat{\rho}^{Q}(t)$ is thus given by
\begin{eqnarray}
\rho^{Q}_{\sigma\sigma}(t) & = & K^{Q,\sigma\sigma}_{a}(t) \, \rho^{Q}_{\sigma\sigma}(0) + K^{Q,\sigma}_{b}(t) \, \rho^{Q}_{\bar{\sigma}\bar{\sigma}}(0) \,\,, \\
\rho^{Q}_{\sigma\bar{\sigma}}(t) & = & K^{Q,\sigma\bar{\sigma}}_{a}(t) \, \rho^{Q}_{\sigma\bar{\sigma}}(0)  \,\, ,
\end{eqnarray}
where we see that the dynamics of coherences ($\rho_{+-} \! = \! \rho^{*}_{-+}$) is decoupled from the dynamics of the populations ($\rho_{++}$ and $\rho_{--}$). This was first noted in Ref.~\onlinecite{Coish_PRB04}. Furthermore, from $\rho_{++}+\rho_{--} \! =\! 1$ we have 
$K^{Q,\sigma}_{b}(t) \!= \! 1 - K^{Q,\bar{\sigma}\bar{\sigma}}_{a}(t)$. 

In an analogous way we derive the expression for time evolution of the two-qubit reduced density matrix in the case of uncorrelated baths:
\begin{align}
\rho_{\sigma_{A}\sigma_{B},{\sigma'}_{A}{\sigma'}_{B}}(t) & =   \!\!\!\!\! \sum_{\xi_{A},\xi_{B},{\xi'}_{A},{\xi'}_{B}} \!\!\! \rho_{\xi_{A}\xi_{B},{\xi'}_{A}{\xi'}_{B}}(0) \nonumber \\
& \!\!\! \times K^{A}_{\sigma_{A}\xi_{A},{\xi'}_{A}{\sigma'}_{A}}(t)  K^{B}_{\sigma_{B}\xi_{B},{\xi'}_{B}{\sigma'}_{B}}(t)   \,\, ,  \label{eq:rhoKdef}
\end{align}
and using Eq.~(\ref{eq:Kab}) we arrive at specific form of $R$ function, in which the four terms on the RHS of Eq.~(\ref{eq:Rabcd4}) are respectively given by $K^{A,\alpha\gamma}_{a}K^{B,\beta\delta}_{a}$, $K^{A,\alpha\gamma}_{a}K^{B,\delta}_{b}$, $K^{A,\alpha}_{b}K^{B,\beta\delta}_{a}$, and $K^{A,\alpha}_{b}K^{B,\beta}_{b}$. 

Using the above formulas  we obtain the diagonal elements of the two-qubit density matrix,
\begin{align}
\rho_{\sigma\xi,\sigma\xi}(t) & = K^{A,\sigma\sigma}_{a} K^{B,\xi\xi}_{a}\rho_{\sigma\xi,\sigma\xi}(0) 
                                + K^{A,\sigma\sigma}_{a} K^{B,\xi}_{b}\rho_{\sigma\bar{\xi},\sigma\bar{\xi}}(0) \nonumber \\
                              & + K^{A,\sigma}_{b} K^{B,\xi\xi}_{a}\rho_{\bar{\sigma}\xi,\bar{\sigma}\xi}(0) 
                                + K^{A,\sigma}_{b} K^{B,\xi}_{b}\rho_{\bar{\sigma}\bar{\xi},\bar{\sigma}\bar{\xi}}(0) \,\, , \label{eq:rho-diag}
\end{align}
where the time-dependence of $K(t)$ functions has been suppressed for clarity of notation. For off-diagonal elements corresponding to coherences between the states differing by a single spin-flip, we have
\begin{align}
\rho_{\sigma\xi,\sigma\bar{\xi}}(t) & = K^{A,\sigma\sigma}_{a}K^{B,\xi\bar{\xi}}_{a}\rho_{\sigma\xi,\sigma\bar{\xi}}(0) + 
K^{A,\sigma}_{b}K^{B,\xi\bar{\xi}}_{a}\rho_{\bar{\sigma}\xi,\bar{\sigma}\xi}(0) \,\, , \nonumber\\
\rho_{\sigma\xi,\bar{\sigma}\xi}(t) & = K^{A,\sigma\bar{\sigma}}_{a}K^{B,\xi\xi}_{a}\rho_{\sigma\xi,\bar{\sigma}\xi}(0) + 
K^{A,\sigma\bar{\sigma}}_{a}K^{B,\xi}_{b}\rho_{\sigma\bar{\xi},\bar{\sigma}\bar{\xi}}(0) \,\, , \label{eq:rho_off_single}
\end{align}
and for coherences between the states differing by two spin flips we have
\beq
\rho_{\sigma\xi,\bar{\sigma}\bar{\xi}}(t)  = K^{A,\sigma\bar{\sigma}}_{a}K^{B,\xi\bar{\xi}}_{a}\rho_{\sigma\xi,\bar{\sigma}\bar{\xi}}(0) \,\, .\label{eq:rho_off_double}
\eeq

A simple calculation shows that the electron spin structure of Eqs.~(\ref{eq:rho-diag})--(\ref{eq:rho_off_double}) holds also in the general case described by Eqs.~(\ref{eq:evo_general})  and (\ref{eq:Rabcd4}), in which the interbath correlations are allowed. 
This has an important consequence for entanglement dynamics. Below we consider initial states of the qubits described by a density matrix diagonal in the basis of Bell states. 
This matrix in the product basis $\ket{\sigma_{A}\sigma_{B}}$ has non-zero populations, and non-zero coherences between the states differing by two spin flips (non-zero values only on the diagonal and the antidiagonal, i.e.~it is of the $X$ form). From the above equations of motion we see that this $X$ form is preserved at all times. 

\subsection{General formalism in the spin echo case} \label{sec:evo_SE}
Since the spin echo protocol leads by itself to removal of inhomogeneous broadening, there is little incentive to consider narrowed bath states, and we focus now on thermal and uncorrelated baths.

In the protocol, a $\pi$ pulse (taken as a perfect pulse along the $x$ axis here) is applied to each qubit at time $\tau$, while the coherence measurement is performed at time $2\tau$. The evolution operator $\hat{U}_{Q}$ in Eq.~(\ref{eq:Kdef}) is then given by
\beq
\hat{U}^{\text{SE}}_{Q}(2\tau) = \text{e}^{-i\hat{H}_{Q}\tau}(-i\hat{\sigma}_{x})\text{e}^{-i\hat{H}_{Q}\tau} \,\, ,
\eeq
and the $K(2\tau)$ functions derived using the above operator describe the evolution of the two-qubit density matrix from the initial state to the final state at time $2\tau$, at which the coherence revival is expected to be maximal. With these functions we can calculate the $2\tau$-dependence (i.e.~the dependence on the total echo sequence time) of the reduced density matrix of the qubits. Alternatively, in order to visualize how the echo signal appears as the time of final measurement is varied, we can use the evolution operator 
\beq
\hat{U}^{\text{SE}}_{Q}(\tau_{1},\tau_{2}) = \text{e}^{-i\hat{H}_{Q}\tau_{2}}(-i\hat{\sigma}_{x})\text{e}^{-i\hat{H}_{Q}\tau_{1}} \,\, ,
\eeq
and investigate the dependence of the reduced density matrix on $\tau_{1}$ and $\tau_{2}$. Of course we expect to see a partial recovery of the signal for $\tau_{1} \! \approx \! \tau_{2}$.

Depending on the way in which we want to look at the results of the echo protocol, we obtain $K_{\sigma\xi,\xi'\sigma'}$ functions with either $2\tau$ (the total time of sequence with $\pi$ pulse at $\tau$) or $\tau_{1}$ and $\tau_{2}$ as their arguments. The structure of the functions is the same in both cases as long as $\tau_{2} \! > \! 0$. However, it is richer than the structure of $K(t)$ functions discussed previously for the free evolution case. Now we have
\begin{align}
K_{\sigma\xi,\xi'\sigma'} & \equiv \delta_{\sigma\bar{\xi}}\delta_{\sigma'\bar{\xi}'}K^{\sigma\sigma'}_{a} 
 + \delta_{\sigma\sigma'}\delta_{\sigma\bar{\xi}}\delta_{\sigma\bar{\xi}'}K^{\sigma}_{b} \nonumber \\
& + \delta_{\sigma\sigma'}\delta_{\sigma\xi}\delta_{\sigma\xi'}K^{\sigma}_{c} 
 + \delta_{\sigma\bar{\sigma}'}\delta_{\sigma\xi}\delta_{\sigma\bar{\xi}'} K^{\sigma}_{d} \,\, ,\label{eq:KSE}
\end{align}
where
\begin{widetext}
\begin{align}
K^{\sigma\sigma'}_{a} & \doteq \sum_{m}\text{Tr}(\hat{P}_{\sigma'\sigma}\hat{U}_{2}^{m,m}\hat{P}_{\sigma\bar{\sigma}}\hat{U}_{1}^{m,m}\hat{P}_{\bar{\sigma}\bar{\sigma}'}\hat{\rho}_{J} (\hat{U}_{1}^{\dagger})^{m,m}\hat{P}_{\bar{\sigma}'\sigma'}(\hat{U}_{2}^{\dagger})^{m,m} )  \,\, , \label{eq:KSEa} \\
K^{\sigma}_{b} & \doteq \sum_{m}\text{Tr}(\hat{P}_{\sigma\sigma}\hat{U}_{2}^{m,m+\sigma}\hat{P}_{\bar{\sigma}\sigma}\hat{U}_{1}^{m+\sigma,m+2\sigma}\hat{P}_{\bar{\sigma}\bar{\sigma}}\hat{\rho}_{J} (\hat{U}_{1}^{\dagger})^{m+2\sigma,m+\sigma}\hat{P}_{\sigma\bar{\sigma}}(\hat{U}_{2}^{\dagger})^{m+\sigma,m} )  \,\, , \label{eq:KSEb} \\
K^{\sigma}_{c} & \doteq \sum_{m}\text{Tr}(\hat{P}_{\sigma\sigma}\hat{U}_{2}^{m,m}\hat{P}_{\sigma\bar{\sigma}}\hat{U}_{1}^{m,m-\sigma}\hat{P}_{\sigma\sigma}\hat{\rho}_{J} (\hat{U}_{1}^{\dagger})^{m-\sigma,m}\hat{P}_{\bar{\sigma}\sigma}(\hat{U}_{2}^{\dagger})^{m,m} ) \nonumber \\ 
& + \sum_{m}\text{Tr}(\hat{P}_{\sigma\sigma}\hat{U}_{2}^{m,m+\sigma}\hat{P}_{\bar{\sigma}\sigma}\hat{U}_{1}^{m+\sigma,m+\sigma}\hat{P}_{\sigma\sigma}\hat{\rho}_{J} (\hat{U}_{1}^{\dagger})^{m+\sigma,m+\sigma}\hat{P}_{\sigma\bar{\sigma}}(\hat{U}_{2}^{\dagger})^{m+\sigma,m} ) \,\, , \label{eq:KSEc} \\
K^{\sigma}_{d} & \doteq \sum_{m}\text{Tr}(\hat{P}_{\bar{\sigma}\sigma}\hat{U}_{2}^{m,m}\hat{P}_{\sigma\bar{\sigma}}\hat{U}_{1}^{m,m-\sigma}\hat{P}_{\sigma\bar{\sigma}}\hat{\rho}_{J} (\hat{U}_{1}^{\dagger})^{m-\sigma,m-\sigma}\hat{P}_{\bar{\sigma}\sigma}(\hat{U}_{2}^{\dagger})^{m-\sigma,m} ) \nonumber \\ 
& + \sum_{m}\text{Tr}(\hat{P}_{\bar{\sigma}\sigma}\hat{U}_{2}^{m,m+\sigma}\hat{P}_{\bar{\sigma}\sigma}\hat{U}_{1}^{m+\sigma,m+\sigma}\hat{P}_{\sigma\bar{\sigma}}\hat{\rho}_{J} (\hat{U}_{1}^{\dagger})^{m+\sigma,m}\hat{P}_{\sigma\bar{\sigma}}(\hat{U}_{2}^{\dagger})^{m,m} ) \,\, , \label{eq:KSEd}
\end{align}
\end{widetext}
where we have defined
$\hat{U}_{1/2}^{n,m} \! \doteq \! \hat{\Pi}_{n}\hat{U}(\tau_{1/2})\hat{\Pi}_{m}$ and $(\hat{U}^{\dagger}_{1/2})^{n,m} \! \doteq \!  \hat{\Pi}_{n}\hat{U}^{\dagger}(\tau_{1/2})\hat{\Pi}_{m}$.
Note that the structure of $\hat{\rho}(2\tau)$ is now more complicated than in the case of free evolution. The diagonal element $\rho_{\sigma\xi,\sigma\xi}(2\tau)$ still depends only on the initial values of all the diagonal elements,
\begin{align}
\rho_{\sigma\xi,\sigma\xi}(2\tau) & = K^{A,\sigma}_{b} K^{B,\xi}_{b}\rho_{\sigma\xi,\sigma\xi}(0) \nonumber \\
& + \left(K^{A,\sigma}_{b} K^{B,\xi\xi}_{a} + K^{A,\sigma}_{b}K^{B,\xi}_{d}\right) \rho_{\sigma\bar{\xi},\sigma\bar{\xi}}(0) \nonumber \\
& + \left(K^{A,\sigma\sigma}_{a} K^{B,\xi}_{b} + K^{A,\sigma}_{d}K^{B,\xi}_{b}\right)  \rho_{\bar{\sigma}\xi,\bar{\sigma}\xi}(0) \nonumber \\
& + \left(K^{A,\sigma\sigma}_{a} K^{B,\xi\xi}_{a}  + K^{A,\sigma\sigma}_{a}K^{B,\xi}_{d} \right.  \nonumber\\ 
&   \left.+K^{A,\sigma}_{d}K^{B,\xi\xi}_{a}+K^{A\sigma}_{d}K^{B,\xi}_{d} \right) \rho_{\bar{\sigma}\bar{\xi},\bar{\sigma}\bar{\xi}}(0) \,\, , \label{eq:rho-diag-echo}
\end{align}
but the behavior of coherences is different. The coherences between the states differing by a single spin flip depend now on the initial values of all the coherences of this kind, and the same holds for the coherences between the states differing by two spin flips -- but the two groups of coherences are mutually independent. As we will see below, the decay of entanglement of Bell states is determined by dynamics of occupations and two-spin-flip coherence, so we give here the formula for the latter:
\begin{align}
\rho_{\sigma\xi,\bar{\sigma}\bar{\xi}}(2\tau) & = K^{A,\sigma}_{c}K^{B,\xi}_{c}\rho_{\sigma\xi,\bar{\sigma}\bar{\xi}}(0) + K^{A,\sigma}_{c}K^{B,\xi\bar{\xi}}_{a}\rho_{\sigma\bar{\xi},\bar{\sigma}\xi}(0)  \nonumber\\
& + K^{A,\sigma\bar{\sigma}}_{a}K^{B,\xi}_{c}\rho_{\bar{\sigma}\xi,\sigma\bar{\xi}}(0) + K^{A,\sigma\bar{\sigma}}_{a}K^{B,\xi\bar{\xi}}_{a}\rho_{\bar{\sigma}\bar{\xi},\sigma\xi}(0) \,\, .
\end{align} 
Although the exact dynamics is more complicated than in the free evolution case, a state initialized in an $X$ form again retains this structure throughout the evolution. 

\subsection{Approximation schemes for evolution functions at lowest magnetic fields: uniform coupling approach}  \label{sec:UC}
In order to make use of the above schemes of calculation of the two-qubit density matrix, we have to choose an approximate method of calculation of $K(t)$ functions (or $R(t)$ functions in the case of correlated baths). At lowest magnetic fields we will use the uniform coupling (UC) approach, in which all $A_{k}$ couplings are assumed to be the same, all equal to $\mathcal{A}/N$. Furthermore, we assume that all the nuclei have the same Zeeman splitting $\omega$, so that we consider a single nuclear macrospin $\mathbf{\hat{J}} = \sum_{k=1}^{N}\mathbf{\hat{J}}_{k}$. 
As discussed in Sec.~\ref{sec:moderate}, this approximation is expected to be justified at low fields, at which the coherence decays on timescale $t\ll \! N/\mathcal{A}$, $\text{min}_{k,l}[1/(\omega_{k}-\omega_{l})]$.

When we replace the nuclear operators by a collective angular momentum operator $\hat{\mathbf{J}}$, we can work in the basis of collective nuclear spin states $\ket{\zeta,j,m}$, for which $\hat{J}^{2}\ket{\zeta,j,m} \! = \! j(j+1) \ket{\zeta,j,m}$ and $\hat{J}^{z}\ket{\zeta,j,m} \! =\! m \ket{\zeta,j,m}$, and where $\zeta$ is the quantum number accounting for many ways in which $N$ spins can be added to a total spin $j$.
The Hamiltonian for a single dot
\beq
\hat{H}_{\text{UC}} = \Omega \hat{S}^{z} + \omega \hat{J}^{z} + \frac{\mathcal{A}}{N}\mathbf{\hat{S}}\cdot\mathbf{\hat{J}} 
\eeq
is then coupling only pairs of states:
\begin{align}
\text{e}^{-i\hat{H}_{\text{UC}}t} \ket{\sigma; \zeta,j,m}  & \equiv a_{jm\sigma}(t) \ket{\sigma; \zeta,j,m} \nonumber \\
& + b_{jm\sigma}(t) \ket{\bar{\sigma}; \zeta,j,m+\sigma/2} \,\, ,  \label{eq:ab}
\end{align}
without leaving the subspace of fixed $j$ and $\zeta$. As explained in Sec.~\ref{sec:bath}, the nuclear density matrix is diagonal in the basis of eigenstates of $\hat{J}^{z}$, so that it can be written as
\beq
\hat{\rho}_{J} = \frac{1}{Z} \sum_{j,m,\zeta} p_{m} \ket{\zeta,j,m}\bra{\zeta,j,m}
\eeq
where $p_{m}$ are the appropriate weights (i.e.~$p_{m}\! = \! 1$ for the thermal state and $p_{m} \! = \! \delta_{mm_{0}}$ for narrowed state), and $Z$ is the statistical sum. Since the Hamiltonian is diagonal in $\zeta$ quantum numbers, the summation over them can be performed at this point:
\beq
\hat{\rho}_{J} = \frac{1}{Z} \sum_{jm} n_{j} p_{m} \ket{j,m}\bra{j,m} \,\, , \label{eq:rhonj}
\eeq
where $n_{j}$ is the number of subspaces with given $j$ (see Appendix \ref{app:UC}).

With the above nuclear density matrix, using Eq.~(\ref{eq:ab}) and the results given in the previous Section, we arrive at $K$ functions in the case of free evolution:
\begin{align}
K_{a}^{Q,\sigma\sigma'}(t) & = \frac{1}{Z}\sum_{jm}n_{j} a^{Q}_{jm\sigma}(t) \big( a^{Q}_{jm\sigma'}(t) \big)^{*} \,\, , \label{eq:KaUCmain} \\
K_{b}^{Q,\sigma}(t) & =  \frac{1}{Z}\sum_{jm}n_{j} |b^{Q}_{jm\bar{\sigma}}(t)|^2 \,\,  \label{eq:KbUCmain}.
\end{align}

The calculation for the case of correlated baths described by a density matrix from Eq.~(\ref{eq:rhocorr}), in which in the UC approximation we have $h^{z}_{Q} = m_{Q}\mathcal{A}/N_{Q}$, yields
\begin{align}
& R_{\alpha \beta \gamma \delta}^{\alpha^{\prime} \beta^{\prime} \gamma^{\prime} \delta^{\prime}} \! (t)   = \frac{1}{Z} \sum_{ m_A - m_B = \Delta m } \,\,\,\, \sum_{j_A, \, j_B} n_{j_A} n_{j_B} \nonumber \\
  &\Big(~~ a_{j_A m_A \alpha^{\prime}}  \Big.
          a_{j_B m_B \beta^{\prime}}
          a_{j_A m_A \gamma^{\prime}}^*
          a_{j_B m_B \delta^{\prime}}^*
          \delta_{\alpha \alpha^{\prime}} 
          \delta_{\beta \beta^{\prime}} 
          \delta_{\gamma \gamma^{\prime}} 
          \delta_{\delta \delta^{\prime}} \nonumber \\
      &+  a_{j_A m_A \alpha^{\prime}}
          a_{j_A m_A \gamma^{\prime}}^*
          |b_{j_B m_B \beta^{\prime}}|^2
          \delta_{\alpha \alpha^{\prime}} 
          \delta_{\bar{\beta} \beta^{\prime}} 
          \delta_{\gamma \gamma^{\prime}} 
          \delta_{\beta \delta}
          \delta_{\bar{\beta} \delta^{\prime}}  \nonumber \\
       &+ a_{j_B m_B \beta^{\prime}}
          a_{j_B m_B \delta^{\prime}}^*
          |b_{j_A m_A \alpha^{\prime}}|^2
          \delta_{\alpha \bar \alpha^{\prime}} 
          \delta_{\beta \beta^{\prime}} 
          \delta_{\delta \delta^{\prime}}
          \delta_{\alpha \gamma}
          \delta_{\bar{\alpha} \gamma^{\prime} } \nonumber \\
       &+ |b_{j_A m_A \alpha^{\prime}}|^2
          |b_{j_B m_B \beta^{\prime}}|^2
          \delta_{\alpha \gamma}
          \delta_{\bar \alpha \alpha^{\prime}} 
          \delta_{\bar{\alpha} \gamma'} \delta_{\beta \delta} \delta_{\bar{\beta} \beta'} \delta_{\bar{\beta} \delta'} \Big. \Big) \,\, , \label{eq:R_UC_corr}
\end{align}
where the time-dependence of $a$ and $b$ functions, and the superscripts $Q\! = \! A$, $B$ have been suppressed for clarity. The structure of Eq.~(\ref{eq:Rabcd4}) is of course reproduced here.

As mentioned before, in the echo case we focus on uncorrelated thermal baths. Defining $a_{j,m,\sigma}^{(k)} \! \doteq \! a_{j,m,\sigma}(\tau_{k})$ and $b_{j,m,\sigma}^{(k)} \! \doteq \! b_{j,m,\sigma}(\tau_{k})$ (with $k\! = \! 1, \, 2$), we have the single-qubit evolution functions:
\begin{align}
K_{a}^{\sigma\sigma'}(\Delta t) & = \sum_{jm} n_j ~ a_{j, m, \bar{\sigma}}^{(1)} ~ a_{j, m, \sigma}^{(2)} ~ a_{j, m, \bar{\sigma}^{\prime}}^{(1)*} ~ a_{j, m, \sigma^{\prime}}^{(2)*} \,\, ,  \\
K_{b}^{\sigma}(\Delta t) & = \sum_{jm} n_j \Big( |a_{j, m, \sigma}^{(1)}|^2 ~ |b_{j, m, \bar{\sigma}}^{(2)}|^2 \nonumber \\
                  & + |b_{j, m, \sigma}^{(1)}|^2 ~ |a_{j, m+\sigma, \sigma}^{(2)}|^2 \Big) \,\, , \\
K_{c}^{\sigma}(\Delta t) & = \sum_{jm} n_j \Big( a_{j, m, \sigma}^{(1)} ~ b_{j, m, \bar{\sigma}}^{(2)} ~ b_{j,m, \bar{\sigma}}^{(1)*} ~ a_{j, m+\bar{\sigma}, \bar{\sigma}}^{(2)*} \nonumber \\
                  & + b_{j, m, \sigma}^{(1)} ~ a_{j, m+\sigma, \sigma}^{(2)} ~ a_{j, m, \bar{\sigma}}^{(1)*} ~ b_{j, m, \sigma}^{(2)*} \Big) \,\, \\
K_{d}^{\sigma}(\Delta t) & = \sum_{jm} n_j ~ |b_{j, m, \bar{\sigma}}^{(1)}|^2 ~ |b_{j, m+\bar{\sigma}, \bar{\sigma}}^{(2)}|^2 \,\, ,
\end{align}
where $\Delta t \! \doteq \! \tau_{1}+\tau_{2}$, and the peak of the echo signal is obtained for $\tau_{1} \! = \! \tau_{2} \! = \! \Delta t/2$.

The full formulas for $a_{jm\sigma}$ and $b_{jm\sigma}$ functions are given in Appendix \ref{app:UC}. For the qualitative discussion here we only need to note that for qubit splitting $\Omega\! \ll \! \so$ we have $|a_{jm\sigma}|\!  \approx \! 1$ and $|b_{jm\sigma}| \! \propto \! \so/\Omega$. Thus, at large fields, all the $K_{i}$ functions containing at least one $b$ function are suppressed. This simply means that at large $\Omega$ the nuclear bath leads to pure dephasing of the qubit: only the $K_{a}^{+-}$ functions, describing the decay of coherences which are non-zero in the initial state, are relevant then. This observation is of course general, in the UC approximation we simply can give closed formulas for the $K_{i}$ functions.

\subsection{Effective Hamiltonian theory at moderate magnetic fields} \label{sec:RDT}
As mentioned in Sec.~\ref{sec:moderate}, in NFID and echo experiments, with increasing $\Omega$ the characteristic timescale at which coherence (and entanglement) is substantial becomes longer, and the conditions for $t$ used in the previous Section might not be fulfilled. The case of spin echo in QDs with mutliple nuclear species is of special importance here, since non-trivial echo dynamics occurs then for $t\! > \! \text{min}_{k,l}[1/(\omega_{k}-\omega_{l})]$ while $t\! \ll \! N/\MA$ is still fulfilled.\cite{Cywinski_PRB09,Bluhm_NP10,Neder_PRB11} For unpolarized bath both NFID and echo can be then treated with theory\cite{Cywinski_PRL09,Cywinski_PRB09} using a second-order (in $\Vff$) effective Hamiltonian
\beq
\tilde{H}^{(2)} = \hat{S}^{z}\sum_{k,l}\frac{A_{k}A_{l}}{4\Omega}( \hat{J}^{+}_{k}\hat{J}^{-}_{l} +\hat{J}^{+}_{l}\hat{J}^{-}_{k} ) \,\, ,
\eeq
which can be applicable when $\Omega \! \gg \! \so$.
Using this pure dephasing approximation it is possible to use diagrammatic linked cluster expansion, and to perfom a resummation of the so-called ring diagrams to derive a closed formula for $K^{\sigma\bar{\sigma}}_{a}(t)$ (which is the only non-trivial evolution function in this case). Using various degrees of coarse-graining of distribution of hf couplings, one can see that for $t\! \ll \! N/\MA$, in this approach we can indeed assume all the $A_{k}$ to be equal. The result for NFID obtained in this way agrees\cite{Barnes_PRB11} with the UC calculation for $\Omega \! \gg \! \so$ (only the small-asmplitude oscillations of $K^{\sigma\bar{\sigma}}_{a}(t)$ are missing, see below). More importantly, for spin echo, for $t\! \ll \! N/\MA$ this approach gives an analytical expression for coherence decay,\cite{Cywinski_PRL09,Cywinski_PRB09} 
which agrees with experiments on GaAs quantum dots performed in the relevant $B$ range.\cite{Bluhm_NP10} 
The details of this ring diagram theory (RDT) are given in Ref.~\onlinecite{Cywinski_PRB09} 
(see also an alternative derivation for the echo case in Ref.~\onlinecite{Neder_PRB11}, where the semiclassical nature of the solution is discussed). 

\section{Quantification of entanglement of two spins} \label{sec:quantification}
We assume that the two qubits are intialized in one of Bell states, $\ket{\Phi_{\pm}} \! = \! (\ket{\uparrow\uparrow} \pm \ket{\downarrow\downarrow})/\sqrt{2}$ or $\ket{\Psi_{\pm}} \! = \! (\ket{\uparrow\downarrow} \pm \ket{\downarrow\uparrow})/\sqrt{2}$, or in a Werner state:
\beq
\hat{\rho}_{W} = \frac{1-p}{4}\mathds{1} + p \ket{\Psi_{-}}\bra{\Psi_{-}} \,\, ,  \label{eq:Werner}
\eeq
where $p \! \in \! [0,1]$ can be interpreted as the probability that the entangled $\ket{\Psi_{-}}$ state was indeed prepared. This state represents the simplest example of an imperfectly prepared entangled (for $p\! > \! 1/3$) state of two qubits. We will also see that under the influence of the nuclear bath, a Bell state evolves into a Werner-like state (with additionally decreased coherence) at long times.

Due to the interaction with the environment, even if the initialized two-electron state is pure, the reduced density matrix of the electron spins will become mixed as time progresses. In order to quantify the amount of entanglement in a mixed state of two qubits, many measures can be adopted.\cite{Plenio_QIC07,Aolita_RPP15} The most commonly used one is concurrence,\cite{Wooters_PRL98} $C(t) \! = \! C(\hat{\rho}(t)) \! \in \! [0,1]$ (where $0$ means that the state is separable, and $1$ means that the state is maximally entangled), which is given by
\beq
C(t) = \text{max}(0,\sqrt{\lambda_{1}}-\sqrt{\lambda_{2}}-\sqrt{\lambda_{3}}-\sqrt{\lambda_{4}}) \,\, ,
\eeq
where $\lambda_{i}$ are eigenvalues of $\hat{\rho}(t)(\hat{\sigma}_{y}\otimes\hat{\sigma}_{y})\hat{\rho}^{*}(t)(\hat{\sigma}_{y}\otimes\hat{\sigma}_{y})$ sorted in descending order. According to discussion from Secs.~\ref{sec:evo_FID} and \ref{sec:evo_SE}, a state initially having the $X$ form (e.g.~one of the Bell states, or the Werner state) maintains this form during the evolution, so it is useful to note that for $X$-states we have:\cite{Yu_QIC07}
\beq
C = 2 \, \text{max}(0,|\rho_{14}|-\sqrt{\rho_{22}\rho_{33}},|\rho_{23}|-\sqrt{\rho_{11}\rho_{44}}) \,\, , \label{eq:CX}
\eeq
where $\rho_{ij}$ are the matrix elements (in the standard $\ket{\sigma_{A}\sigma_{B}}$ basis) of $\hat{\rho}(t)$ (for a basis-independent expression for $C(t)$ involving expectation values of spherical tensor operators see Ref.~\onlinecite{Szankowski_arXiv14}). 

Let us note now that when considering the initial Bell state at very high magnetic fields, at which only pure dephasing occurs, we simply have $C(t)\! \approx \! 2|\rho_{ab}(t)|$, where $\rho_{ab}(t)$ is the initially non-zero coherence. In the considered here case of uncoupled qubits the calculation of entanglement decay amounts then to multiplication of known high-field results for single-qubit decoherence. Here we focus mostly on low and moderate $B$ fields, at which $C(t)$ exhibits features not present in single-qubit coherence decay.

Comparing the theoretical predictions for $C(t)$  with experiment requires performing a full tomographic reconstruction of $\hat{\rho}(t)$ (see e.g.~Ref.~\onlinecite{Shulman_Science12} for experimental example with DQDs, and Ref.~\onlinecite{Rohling_PRB13} for theoretical proposal of tomography scheme for DQDs taking into account the limitations specific to experiments on gated QDs). Since the two-qubit tomography requires at least 15 different measurement settings, simpler ways to quantify entanglement  are clearly of interest, even if they apply only to certain kinds of entangled states. For example, entanglement witnesses\cite{Guhne_PR09} are observables which have negative expectation value only when the state is entangled. Note that not every entangled state is detected by a given witness, and in fact the most natural construction of a witness begins with assuming certain form of a mixed entangled state which one aims to detect.\cite{Guhne_PRA02} For electrically controlled DQDs, the projection on a 
singlet state, $\hat{P}_{S}$,  is a natural two-qubit measurement operator,\cite{Hanson_RMP07} and consequently we have a witness $\hat{w}_{S}\! \equiv \! 1/2-\hat{P}_{S}$. It should be noted that a decay of an entangled two-spin singlet state in a DQD was monitored through observation of $\mean{\hat{P}_{S}(t)}$ in the first paper on coherent control of singlet-triplet qubit.\cite{Petta_Science05}
Witnesses corresponding to other Bell states can be obtained by first subjecting the two qubits to gate operations converting a given Bell state into a singlet. 

The presence of a specific kind of entanglement can also be verified by using the given state to perform a certain task and achieving a result which is proven to be impossible to obtain using separable states. For two qubits, a natural example of such a task is quantum teleportation.\cite{Bennett_PRL93,Horodecki_PRA99,Verstraete_PRL03,vanEnk_PRA07} With an entangled state $\hat{\rho}_{AB}$ of spins $A$ and $B$ (located in possibly distant QDs) established, and with a third dot $C$ located close to dot $A$ (in order to allow for two-qubit operations on $A$ and $C$), a quantum state of spin $C$ can be imprinted on spin $B$ with the use of two-qubit gate on $A$ and $C$, and with single-qubit operations and projective measurements. We assume that a Bell states of spins $A$ and $B$ is initialized, and then at a later time $t$ a pure state $\ket{\phi_{C}}$ is created, and teleportation protocol is carried out on timescale negligible compared to timescales of qubits' dynamics. 
Assuming that all the gate operations are error-free, the fidelity of the state teleported at time $t$,
\beq
F_{\phi}(t) = \bra{\phi_{C}}\text{Tr}_{A} \big( \hat{\rho}_{AB}(t) \big) \ket{\phi_{C}} \,\, ,
\eeq
is determined by the density matrix $\hat{\rho}_{AB}(t)$ at the moment of teleportation. The key observation\cite{Popescu_PRL94,Massar_PRL95} is that the fidelity averaged over all $\ket{\phi_{C}}$ states, $\bar{F}(t)$, cannot exceed $2/3$ for separable $\hat{\rho}_{AB}(t)$, and consequently, the  observation of $\bar{F}\! > \! 2/3$ is a proof of entanglement of spins $A$ and $B$. This procedure of entanglement verification becomes practical when one realizes that instead of averaging over all the $\ket{\phi_{C}}$ states, it is in fact enough to average over six mutually unbiased basis states,\cite{vanEnk_PRA07,Pfaff_Science14} e.g.~$\ket{\pm X}$, $\ket{\pm Y}$, $\ket{\pm Z}$.

\section{Entanglement dynamics due to coupling to thermal and partially narrowed baths} \label{sec:thermal}
When the experimental situation corresponds to averaging of free evolution of qubits over the thermal equilibrium (or partially narrowed) nuclear density matrices, we can, as discussed in Sec.~\ref{sec:decoherence_basics}, use the QSBA as long as we do not consider evolution times much longer than $T_{2}^{*}$. 
This means that we can simply use the UC approach from Sec.~\ref{sec:UC}. The UC solution for free evolution of a spin interacting with a thermal nuclear bath is well known,\cite{Zhang_PRB06,Coish_JAP07} and we include it for the sake of completeness in Appendix \ref{app:UC}, where closed expressions for $K^{\sigma\sigma'}_{a}$ and $K^{\sigma}_{b}$ are given.

For magnetic fields corresponding to $\Omega\! \gg \! \so$, decoherence has mostly the character of pure dephasing: as shown in Appendix \ref{app:UC}, for large bath (i.e.~for large $N$) and at times $t\! \ll \! \Omega/\so^2$ we have
\beq
K^{Q,\sigma\bar{\sigma}}_{a}(t)  \approx \text{e}^{-i\sigma\Omega_{Q}t}\exp \! \Big( \! -(t/T^{*}_{2,Q})^2  \Big) \,\, , \label{eq:Ka_inh}
\eeq
where $Q$ labels the $A$ and $B$ qubits. This is a result that can be easily derived by completely neglecting the $\Vff$ operator and performing a Gaussian average over static $h^{z}$ fields.\cite{Merkulov_PRB02} The $K^{\sigma}_{b}$ functions are then strongly suppressed by large $\Omega$, and consequently $K^{\sigma\sigma}_{a} \! = \! 1 - K^{\bar{\sigma}}_{b} \! \approx \! 1$. The exact form of the transient dynamics of $K^{\sigma}_{b}$ matters little for $\Omega \! \gg \! \so$, but it is important to note that at $t\! > \! T^{*}_{2,Q}$ we obtain 
\beq
K_{b}^{Q,\sigma}(t>T_{2,Q}^{*}) \approx \frac{\sigma^{2}_{Q}}{\Omega^{2}} = \frac{2}{(\Omega T^{*}_{2,Q})^2} \equiv \frac{2}{\tilde{\Omega}^{2}_{Q}}  \,\, , \label{eq:Kb-inh}
\eeq
where $\tilde{\Omega}_{Q} \! \doteq \! \Omega T^{*}_{2,Q}$ is the dimensionless Zeeman splitting of electron in dot $Q$, 
see Appendix \ref{app:UC} for derivation of this formula. 

\begin{figure}
\includegraphics[width=\linewidth]{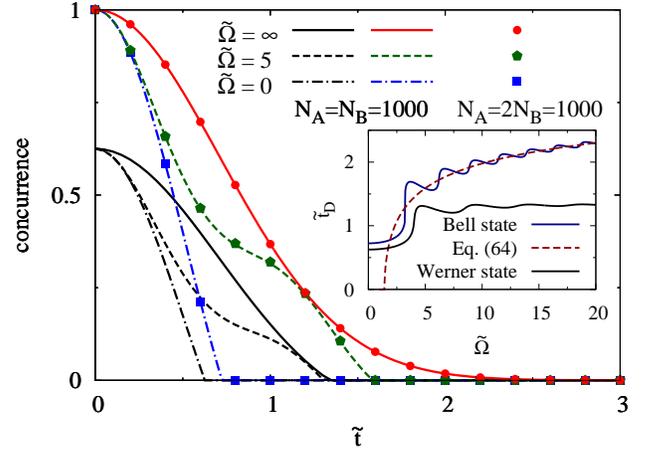}
\caption{(Color online)~Concurrence decay for two spins interacting with uncorrelated thermal baths. The spins are initially in one of the Bell states, and the dynamics of their reduced density matrix is calculated using the UC approach. Time is in units of two-spin $T_{2}^{*}$ defined in Eq.~(\ref{eq:T2star2}), and the dimensionless Zeeman splitting is $\tilde{\Omega} \! \doteq \! \Omega T_{2}^{*}$. The result for $\tilde \Omega \! = \! \infty$ (which in fact is a value large enough for the changes in $C(\tilde t)$ upon increasing it to be invisible in the Figure) is the same as the result of calculation using the pure dephasing approximation from Eq.~(\ref{eq:Ka_inh}). Lines and symbols correspond to two distinct sets of $N_{A}$ and $N_{B}$. The agreement of the results corresponding to the same values of $\tilde{\Omega}$ illustrates the universal character of the $C(\tilde{t})$ behavior. 
Black lines show results for the initial Werner state with $p\! =\! 3/4$.
In the inset we show the dependence of the ESD time $\tilde{t}_{\text{D}} \! \doteq \! t_{D}/T_{2}^{*}$ on $\tilde{\Omega}$: blue solid line is the exact result for Bell states, dashed line is the approximate large-field result from Eq.~(\ref{eq:tESDuniv}), and black solid line is the exact result for Werner state with $p\! =\! 3/4$.} \label{fig:C_thermal}
\end{figure} 

Using Eq.~(\ref{eq:CX}) we immediately see that at high fields for Bell states we have $C(t) \! \approx\! 2|\rho_{ab}(t)|$ (where $\rho_{ab}(t)$ is the non-zero coherence present in a given Bell state) up to entanglement sudden death time $t_{\text{D}}$ at which $|\rho_{ab}| \! = \! \sqrt{\rho_{cc}\rho_{dd}}$ (where all the indices $a$, $b$, $c$, $d$ are distinct):
\beq
C(t < t_{\text{D}}) \approx \exp \! \left( \! -\left( \frac{t}{T_{2}^{*}} \right )^2 \right) \,\, ,  \label{eq:Cinh}
\eeq
where the two-dot $T_{2}^{*}$ time is given by
\beq
\frac{1}{(T_{2}^{*})^2} = \frac{1}{(T_{2,A}^{*})^2} + \frac{1}{(T_{2,B}^{*})^2} \,\, .  \label{eq:T2star2}
\eeq
In order to obtain $t_{\text{D}}$ we have to look at the diagonal elements of $\hat{\rho}(t)$ which initially were equal to zero. 
In high fields, $\tilde{\Omega}_{Q} > 1$, for long times, $t>T_2^*$, using Eqs.~(\ref{eq:rho-diag}) and (\ref{eq:Kb-inh}) and assuming that the initial state was $\ket{\Phi_{\pm}}$ (analogous reasoning applies to any Bell state), we have
\begin{align}
\rho_{22}(t)  & = K^{A,++}_{a}K^{B,-}_{b}\rho_{11}(0) + K^{A,+}_{b}K^{B,--}_{a} \rho_{44}(0) \,\, , \nonumber \\
& = \frac{1}{2}\bigg( \! \left( 1-K^{A,-}_{b} \right)K^{B,-}_{b} + K^{A,+}_{b} \left(1-K^{B,+}_{b}\right) \! \bigg) \,\, , \nonumber\\
& \approx \frac{1}{\tilde{\Omega}^{2}_{A}} + \frac{1}{\tilde{\Omega}^{2}_{B}} \,\, .
\end{align}
The same equation holds for $\rho_{33}(t)$, and solving for $|\rho_{14}(t_{\text{D}})| \! = \! \sqrt{\rho_{22}(t_{\text{D}})\rho_{33}(t_{\text{D}})}$,
where for $\rho_{14} (t) = K_a^{A,+-} K_a^{B,+-}$ we use high-field approximation, Eq.~(\ref{eq:Ka_inh}), and assuming $\Omega_{A}\! =\! \Omega_{B} \! =\! \Omega$
we obtain
\beq
t_{\text{D}} = T_{2}^{*}\sqrt{2\ln\Omega T_{2}^{*}/\sqrt{2}} \,\, . \label{eq:tESD}
\eeq
We see now that if we introduce dimensionless qubit splitting $\tilde{\Omega} \! \doteq \! \Omega T_{2}^{*}$, 
and dimensionless time $\tilde{t} \! \doteq \! t/T_{2}^{*}$, the expressions for both the concurrence and the ESD time become 
\begin{align}
C(t<t_{\text{D}}) & \approx \text{e}^{-\tilde{t}^2} \,\, , \label{eq:Cuniv} \\
\tilde{t}_{\text{D}} & = \sqrt{2\ln \tilde{\Omega}/\sqrt{2}} \,\, . \label{eq:tESDuniv}
\end{align}
Above discussion suggests that this universality is expected at $\tilde{\Omega} \! \gg \! 1$, but the numerical calculations withing the UC model show that it holds at all magnetic fields, with $N \! \gg \! 1$ being the only requirement. In Fig.~\ref{fig:C_thermal} we present the results for $C(\tilde{t})$ at $\tilde{\Omega} \! = \! 0$, $5$, and $\infty$ (in practice a value large enough so that increasing it does not change the result in a visible manner), calculated using either symmetric ($N_{A} \! = \! N_{B}$) or strongly asymmetric ($N_{A} \!= \! 2N_{B}$) dots. While only the large-$\tilde{\Omega}$ result is in quantitative agreement with Eq.~(\ref{eq:Cuniv}), the fact that $C(\tilde{t})$ depends only on $\tilde{\Omega}$ is true for any $\tilde{\Omega}$. As shown in Fig.~\ref{fig:C_thermal}, for $\tilde{\Omega} \! \lesssim \! 1$ the concurrence has an oscillatory component which related to oscillations of $K^{\sigma}_{b}$ functions with frequency $\sim \! \tilde{\Omega}$ (and their appreciable amplitude 
at such low fields), which lead to oscillations of the diagonal elements of the two-qubit density matrix. It is also worth noting that at these low fields the sudden death is a visible effect, i.e.~there is an appreciable discontinuity of derivative of $C(t)$ at the time at which the concurrence becomes zero. The dependence of $\tilde{t}_{\text{D}}$ on $\tilde{\Omega}$ is shown in the inset of Fig.~\ref{fig:C_thermal}. The results for $C(t)$ and $t_{\text{D}}(\Omega)$ in the thermal bath case were previously obtained in Ref.~\onlinecite{Mazurek_PRA14}, where it was suggested that a measurement of $t_{\text{D}}$ can be used for sensing of small magnetic fields, while in Refs.~\onlinecite{Cai_PRL10} and \onlinecite{Tiersch_JPCA14} the possible significance of entanglement dynamics for the operation of chemical magnetometers was discussed.
Here, we have shown that $\tilde{t}_{\text{D}} \! \doteq \! t_{\text{D}}/T_{2}^{*}$ is a function of $\tilde{\Omega}\! \doteq \! \Omega T_{2}^{*}$ \textit{only}. With this insight it is clear that any pair of QDs (even two dots of very different sizes) can be used for such a magnetometry scheme. The high-field measurement of $C(t)$ decay allows for calibration of units (i.e.~it gives the value of $T_{2}^{*}$), and then $\tilde{t}_{\text{D}}$ is a universal function of $\tilde{\Omega}$.

It should be noted that in recent experiments\cite{Brunner_PRL11} on DQDs, in order to separately address each of the qubits (i.e.~perform the single-qubit unitary operations using a.c.~magnetic field), the qubits were exposed to a magnetic field gradient, resulting in $\Omega_{A}$ being slightly different from  $\Omega_{B}$. Writing $\Omega_{A}\! = \! \Omega$ and $\Omega_{B}\! = \! (1+\eta)\Omega$, with $\eta \! \ll \! 1$, we can repeat the above derivation for $\Omega \gg 1 $, and obtain the first-order correction to formula (\ref{eq:tESDuniv}), which we write as $\tilde{t}_{\text{D}} \approx \tilde{t}_{0} +\delta\tilde{t}$ with
\beq
\delta\tilde{t} =  - \eta \frac{(T^{*}_{2,B})^2}{(T^{*}_{2,A})^2+(T^{*}_{2,B})^2} \frac{1}{ \tilde{t}_{0}} \,\, ,
\eeq
where $\tilde{t}_{0}$ corresponds to $\tilde{t}_{D}$ for $\Omega_{A}\! = \! \Omega_{B}$, i.e.~to the value from Eq.~(\ref{eq:tESD}).

A different behavior of $\tilde{t}_{\text{D}}$ arises in the case of initial Werner state from Eq.~(\ref{eq:Werner}). Using Eq.~(\ref{eq:rho-diag}) we obtain (retaining only terms of leading order in $1/\tilde{\Omega}^2$) the $\tilde{t} \! > \! 1$ values of $\rho_{11}\! = \! \rho_{44} \! \approx (1-p)/4 + p^2/\tilde{\Omega}^2$, which together with $|\rho_{23}| \! \approx \frac{p}{2} \text{e}^{-\tilde{t}^2}$ lead to 
\beq
\tilde{t}_{\text{D,W}} = \sqrt{\ln\left[ \left ( \frac{1-p}{2p} + \frac{2}{\tilde{\Omega}^2} \right)^{-1} \right ]} \,\, ,
\eeq
which at large fields, $\tilde{\Omega} \! \gg \! 2\sqrt{p/(1-p)}$, gives an $\tilde{\Omega}$-independent result, $\tilde{t}_{\text{D,W}} \! \approx \! \sqrt{\ln \! \left( 2p/(1-p) \right)}$, visible in the inset of Fig.~\ref{fig:C_thermal} (see also Ref.~\onlinecite{Bellomo_PRA10} for an analogous result obtained for superconducting qubits exposed to $1/f$ phase noise).

Finally, let us underline the basic features of the mixed $\hat{\rho}(t)$ state. The main effect of the nuclear bath is the dephasing of coherences, and the secondary effect at finite magnetic field is a partial redistribution of populations. For an initial Bell state this redistribution increases both initially zero populations by the same amount, and the resulting $\hat{\rho}(t)$ state is a Werner-like state: the initial coherence is suppressed, while a fraction of $\mathds{1}$ state is admixed. This is a general feature of hf-induced dynamics of two-qubit density matrix, which holds for results presented in subsequent sections.

\section{Entanglement dynamics due to coupling to nuclear baths in a strongly narrowed state}  \label{sec:narrowed}
Now we focus on baths in a strongly narrowed state of the nuclear baths, described either by Eqs.~(\ref{eq:rho_narrowed}) and (\ref{eq:rhouncorr}) for uncorrelated baths, or by Eq.~(\ref{eq:rhocorr}) for correlated baths in a DQD. These two cases are considered in Secs.~\ref{sec:Nseparate} and \ref{sec:Ndifference}, respectively.
The fluctuations of $h^{z}_{Q}$ are assumed to be constrained to such a degree that the dephasing due to averaging over a quasistatic spread of $h^{z}_{A,B}$ becomes irrelevant, and other processes (related to averaging over transverse components of Overhauser fields) have to be taken into account. We thus take $\sigma_{\text{n}}$ (the width of relevant distribution of $h^{z}_{Q}$ or $\Delta h^{z}\! \doteq \! h^{z}_{A}-h^{z}_{B}$) to be so small that the timescale $1/\sigma_{\text{n}}$ is much longer than the timescales of decay obtained in what follows.

\subsection{Each bath narrowed separately}  \label{sec:Nseparate}
The calculation of free evolution is performed exactly as in the previous Section, only with the $K_{a/b}(t)$ evolution functions calculated assuming a narrowed state in each of the QDs (see Appendix \ref{app:UC}). For the discussion of the main features of the results we use approximate expressions, valid for $\Omega_{Q} \! \gg \! \sigma_{Q}$, where $\Omega_{Q} \! \doteq \! \Omega + h^{z}_{Q}$ is the total spin splitting in the dot $Q$ (note that spin splitting due to the external $B$ field is again assumed to be the same in both QDs). 
For $\Omega_{Q} \! \gg \! \so$ and on timescales discussed in Appendix \ref{app:UC},  for $h^{z}_{Q}\! = \! 0$ we obtain the smooth coherence decay
\beq
K^{Q,+-}_{a,p=0}(t) \approx \frac{\text{e}^{-i\Omega_{Q} t}}{1+it/\tau_{Q}} = \text{e}^{-i\Omega_{Q} t} \frac{\text{e}^{-i\arctan t/\tau_{Q}}}{\sqrt{1+(t/\tau_{Q})^2}} \,\, ,  \label{eq:NFID_RDT}
\eeq
where
\beq
\tau_{Q} = \frac{4N_{Q}\Omega_{Q}}{\mathcal{A}^{2}} = \frac{1}{2} \Omega_{Q} (T_{2,Q}^{*})^{2} \,\, . \label{eq:tauNFID}
\eeq
The characteristic decay time $\tau_{Q}$ depends now on $\Omega_{Q}$, since the influence of transverse Overhauser fields diminishes with increasing qubit splitting. 

\begin{figure}
\includegraphics[width=\linewidth]{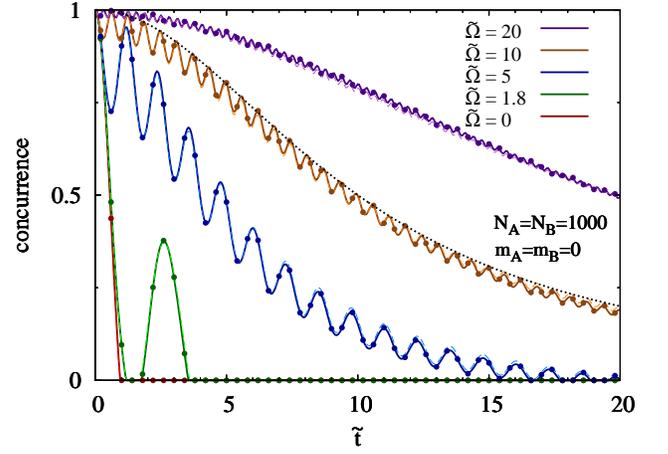}
\caption{(Color online)~Concurrence decay for two electron spins initialized in one of the Bell states interacting with two separate uncorrelated baths in narrowed states (each bath has $h^{z}_{Q} \! = \! 0$). The calculations are performed using the UC approach (so that $h^{z}_{Q} \! = \! 0$ corresponds to $m_{Q}\! = \! 0$). Time is in units of two-spin $T_{2}^{*}$ defined in Eq.~(\ref{eq:T2star2}), and the dimensionless Zeeman splitting is $\tilde{\Omega} \! \doteq \! \Omega T_{2}^{*}$.
Solid lines correspond to the case of two identical QDs ($N_A \! = \! N_B= \! 1000$), 
dashed lines correspond to the case of two strongly assymetric QDs ($N_A \! =2N_B \! = \! 1000$), symbols correspond to the case of two identical QDs consisting of realistic number of nuclear spins ($N_A \! =N_B \! = \! 10^6$), and the dotted line is the calculation in the pure dephasing approximation using Eq.~(\ref{eq:NFID_RDT}) for $\tilde{\Omega}\! = \! 10$.
Note that symbols are in full agreement with solid lines, i.e. the results for two identical QDs are independent of the sizes of these QDs in the domain of applicability fo the UC approach. Dashed lines are very close to the solid ones (the difference between the two is most visible for $\tilde{\Omega}\! = \! 5$)  showing that results obtained for QDs of different sizes, are very similar one to another when expressed in the dimensionless units used here. 
} \label{fig:C_narrowed}
\end{figure}

These results were also obtained using the RDT,\cite{Cywinski_PRL09,Cywinski_PRB09} and they can also be derived by performing a classical average over the transverse components of the Overhauser field.\cite{Cywinski_APPA11,Hung_PRB13} The derivation of this approximate formula from the exact UC solution was sketched in Ref.~\onlinecite{Barnes_PRB11}.

For finite bath polarization and for $t\! \ll \! N_{Q}/\mathcal{A}$ we have\cite{Barnes_PRL12}
\beq
K^{Q,+-}_{a,p}(t) \approx \text{e}^{-i\Omega_{Q} t} \frac{p_{Q}}{p_{Q}\cos (\frac{2Jpt}{\tau_{Q}}) + i p^{2}_{Q,\perp}\sin (\frac{2Jpt}{\tau_{Q}})} \,\, , \label{eq:NFID_RDTp}
\eeq
where $J$ is the nuclear spin (with all the nuclei assumed to have the same $J$), $p_{Q} \! \in \! [0,1]$ is the nuclear polarization, and $p^{2}_{Q,\perp} \! = \! J+1-\mean{(J^{z})^{2}}/J$. Note that for $p_{Q}\! = \! 0$ we have $p^{2}_{\perp} \! = \! J(J+1)/3$, and the $p_{Q} \rightarrow 0$ limit of Eq.~(\ref{eq:NFID_RDTp}) gives Eq.~(\ref{eq:NFID_RDT}).

\begin{figure}
\includegraphics[width=\linewidth]{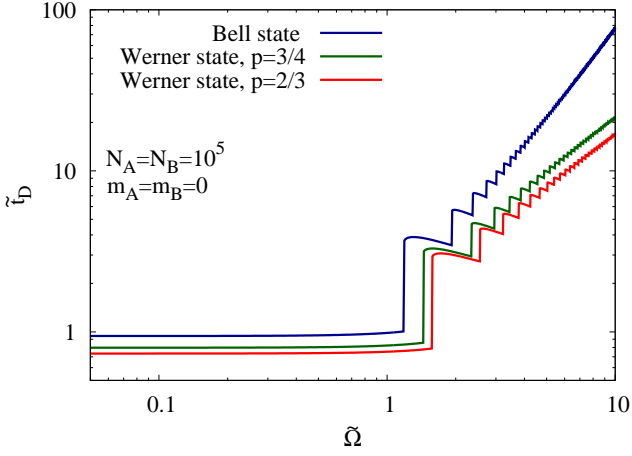}
\caption{(Color online)~Time of final entanglement death for Bell states (the results are the same for all the states) and Werner states (see Eq.~(\ref{eq:Werner})) with $p\! = \! 2/3$ and $3/4$. The nuclear baths for the two dots are uncorrelated, and each is narrowed to the state of $h^{z}\! = \! 0$. Calculation is done within the UC model with $N_{A} \! =\! N_{B} \! = \! 10^{5}$.
Time is in units of two-spin $T_{2}^{*}$ defined in Eq.~(\ref{eq:T2star2}), and the dimensionless Zeeman splitting is $\tilde{\Omega} \! \doteq \! \Omega T_{2}^{*}$.
} \label{fig:ESD_narrowed}
\end{figure}

In the UC solution the above decay functions are modulated by oscillations (with frequency $\approx \tilde\Omega_{Q}$) of amplitude $\approx \! 8/\tilde{\Omega}^{2}_{Q}$, which vanish only for $t \! \gg \! \tau_{Q}$ (see Appendix \ref{app:UC}).
At low $\tilde \Omega_{Q}$ the evolution of diagonal elements of the density matrix is determined by $K^{\sigma}_{b,Q}$ functions, which also exhibit similar oscillations with amplitude $\approx 4/\tilde{\Omega}^{2}_{Q}$, and only for $t \! \gg \! \tau_{Q}$ these oscillations dephase and we have $K^{\sigma}_{b,Q} \! \approx \! 2/\tilde{\Omega}^{2}_{Q}$. These effects can be seen in Fig.~\ref{fig:C_narrowed}, where results for concurrence decay of Bell states at zero bath polarization are shown. At low $\tilde{\Omega}$ these oscillations lead to pronounced effects of entanglement death and revival -- see the result for $\tilde{\Omega} \! = \! 1.8$ for most prominent demonstration. At higher fields they manifest only as a saw-tooth pattern of $\tilde{\Omega}$ dependence of the final ESD time, shown in Fig.~\ref{fig:ESD_narrowed}.

Following the line of reasoning given in the previous Section, we can also derive approximate analytical formulas for ESD time at large magnetic fields. Since their exact forms are much less compact than previously, we only note that  at large $B$ field we have $t_{\text{D}} \! \propto \! B^{2}$ for the initial state being one of the Bell states, and $t_{\text{D}} \! \propto \! B \sqrt{p/(1-p)}$ for the initial state being a Werner state. These asymptotic formulas are confirmed by numerical results shown in Fig.~\ref{fig:ESD_narrowed}.

Finally, let us comment on the fact that in the Figures we are still using the ``universal'' units of $\tilde{\Omega}$ and $\tilde{t}$ introduced in the previous Section. A quick look at the formulas for $K^{\sigma\bar{\sigma}}_{a}$ given above shows than in the fully narrowed case we cannot expect the results for all pairs of dots (e.g.~having $N_{A} \! \neq \! N_{B}$) to collapse on the same curve when using these units. However, we still use them, in order to allow for easy comparison with the results shown previously (and in order to underline the coherence-enhancing effect brought by nuclear bath narrowing). Note also that at short times (when $C(\tilde t) \! \approx \! 1$), for $\tilde \Omega_{A} \! = \! \tilde \Omega_{B}$ we have
$C(\tilde{t}) \approx 1 - 2\tilde{t}^{2}/\tilde{\Omega}^{2}$. At longer times, no universality of results for dots with $T^{*}_{2,A} \! \neq \! T^{*}_{2,B}$ strictly speaking exists. Note however the results shown in the figures are in fact quite representative for pairs of QDs having somewhat different parameters, as proven by the almost complete indistinguishability of solid ($N_{A}\! = \! N_{B}$) and dashed ($N_{A}\! = \! 2N_{B}$) lines in Fig~\ref{fig:C_narrowed}.

\subsection{Narrowing of the Overhauser field gradient -- the case of correlated baths} \label{sec:Ndifference}
When using the correlated state of baths described by Eq.~(\ref{eq:rhocorr}), the first thing worth noticing is the major difference in decoherence of $\ket{\Phi_{\pm}}$ and $\ket{\Psi_{\pm}}$ states in this situation. For the former, the averaging over quasistatic values of $h^{z}_{A}$ and $h^{z}_{B}$ (denoted by $\mean{..}_{z}$) gives
\beq
\rho_{++,--}(t) \propto \mean{ \exp \! \big( \! -i(h^{z}_{A}+h^{z}_{B})t \big) }_{z} \propto \exp \! \Big( \!-(t/T^{*}_{2})^{2} \Big) \,\, ,
\eeq 
where $T_{2}^{*}$ is given by Eq.~(\ref{eq:T2star2}). While the difference of $h^{z}_{A}$ and $h^{z}_{B}$ has diminished fluctuations, the distribution of each of $h^{z}_{A,B}$ is essentially the same as for the thermal bath (with exception of the case of maximal $\Delta h^{z}$ corresponding to two baths fully  polarized in opposite directions). On the other hand, the $\ket{\Psi_{\pm}}$ states are now unaffected by inhomogeneous broadening, since under averaging over $h^{z}_{A,B}$ we have $\rho_{+-,-+}(t) \propto \mean{ \exp(-i\Delta h^{z}t) }_{z} = \exp(-i\Delta h^{z}t) $, and this coherence decays then only due to the influence of the transverse components of the Overhauser fields. All this basically amounts to noticing that the $\ket{\Psi_{\pm}}$ states form a decoherence-free subspace
\cite{Palma_PRSLA96,Lidar_ACP14,Szankowski_arXiv14} with respect to correlated pure-dephasing noise.  

\begin{figure}
\includegraphics[width=\linewidth]{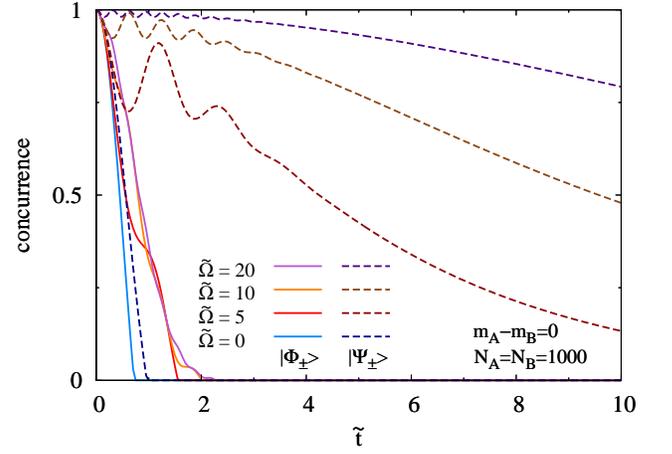}
\caption{(Color online)~Decay of concurrence of $\ket{\Phi_{\pm}}$ and $\ket{\Psi_{\pm}}$ Bell states interacting with correlated nuclear baths in a state of strongly narrowed distribution of $\Delta h^{z} \! \doteq \! h^{z}_{A}-h^{z}_{B}$ (taken to be $\Delta h^{z}\! = \! 0$ here). Calculations are performed using the UC approach.
The $\ket{\Phi_{\pm}}$ states (solid lines) decay just like in the thermal bath case (compare with Fig.~\ref{fig:C_thermal}), while the decay of $\ket{\Psi_{\pm}}$ states (dashed lines) is very similar to the decay observed in the case of separate narrowing of each of the nuclear baths (compare with Fig.~\ref{fig:C_narrowed}), only the fast oscillations of $C(t)$ are absent for $t\! \gtrsim \! T_{2}^{*}$. Time is in units of two-spin $T_{2}^{*}$ defined in Eq.~(\ref{eq:T2star2}), and the dimensionless Zeeman splitting is $\tilde{\Omega} \! \doteq \! \Omega T_{2}^{*}$.
} \label{fig:C_correlated}
\end{figure}

The above observations are illustrated in Fig.~\ref{fig:C_correlated}, where calculations of concurrence obtained in the UC approach using Eqs.~(\ref{eq:evo_general}) and (\ref{eq:R_UC_corr}) for all the Bell states are shown for $\Delta h^{z} \! =\! 0$. The entanglement of $\ket{\Phi_{\pm}}$ states decays as in Fig.~\ref{fig:C_thermal}, while the overall shape of $C(\tilde t)$ dependence for $\ket{\Psi_{\pm}}$ states is in fact very similar to the one from Fig.~\ref{fig:C_narrowed}.

\begin{figure}
\includegraphics[width=\linewidth]{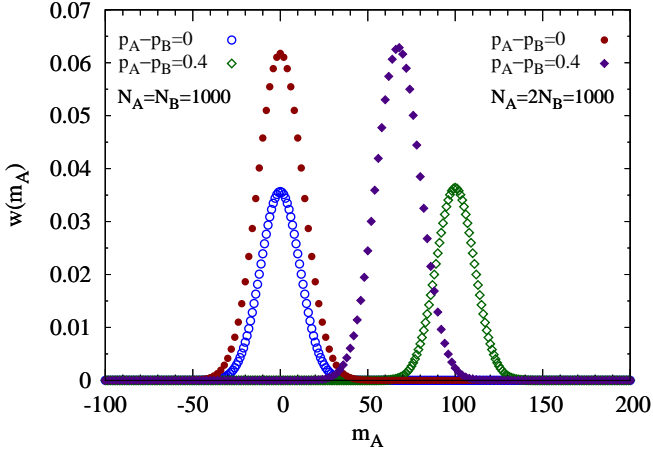}
\caption{(Color online)~Typical distributions of normalized weights for individual $(m_A, m_B)$-pairs for two nuclear spin baths in correlated state 
$\Delta h^z \! \doteq \! h_A^z \! - \!h_B^z \! \Leftrightarrow \! \Delta p \! \doteq \! p_A \! - \! p_B \! = \! \frac{\Delta h^z}{\mathcal{A}J} \! \Rightarrow \!   m_B \! = \! (m_A/N_{A}J - \Delta p)N_{B}J$ in the case of two identical QDs (open symbols) and in the case of two strongly asymmetric QDs (filled symbols). According to the formula given in the main text, the maximum of the distribution occurs at $\bar{m}_{A} \! = \! \Delta p N_{A}N_{B}J/(N_{A}+N_{B})$, which for $\Delta p \! =\! 0.4$ gives $100$ ($66 \frac{2}{3} \! \approx \! 67$) for symmetric (asymmetric) QDs and parameters used here. The normalization is $\sum_{m_A} w(m_A; \Delta m) = 1$.}
\label{fig:CorBaths_weights}
\end{figure}

\begin{figure}
\includegraphics[width=\linewidth]{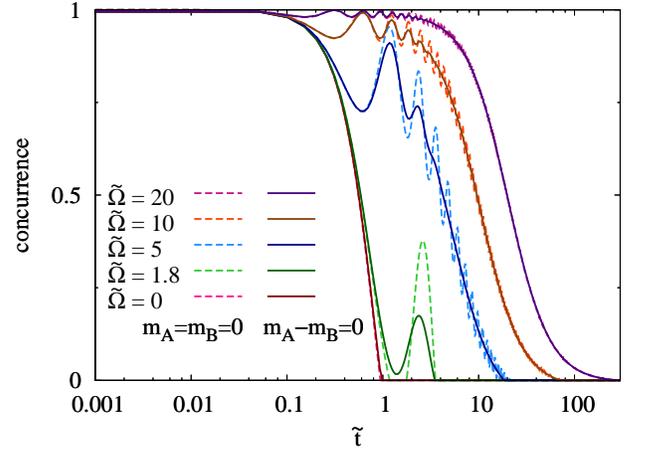}
\caption{(Color online)~Concurrence decay for two electron spins interacting with two uncorrelated and correlated baths in narrowed states. The spins are initially in $|\Psi_{+}\rangle$ or $|\Psi_{-}\rangle$  Bell state and the dynamics of their reduced density matrix is calculated using the UC approach. Dashed lines correspond to the case of separately narrowed baths (calculated for bath sizes $N_A \! = N_B \! = \! 10^5$), while solid lines correspond to the case of correlated state of the baths (calculated for bath sizes $N_A \! = \! N_B \! = \! 1000$) with $\Delta h^{z} \! =\! 0$.
Time is in units of two-spin $T_{2}^{*}$ defined in Eq.~(\ref{eq:T2star2}), and the dimensionless Zeeman splitting is $\tilde{\Omega} \! \doteq \! \Omega T_{2}^{*}$.
} \label{fig:C_narrowed_long}
\end{figure}

The relation between the correlated bath results and the case of uncorrelated baths can be most easily understood when $\Omega \! \gg \! \so$. Then, up to the ESD time, the concurrence for Bell states is approximately proportional to the relevant coherence, which can be related to functions $K^{Q,\sigma\bar{\sigma}}_{a,p_{Q}}$ (where $p_{Q} \! \in \! [-1,1]$ is the polarization of the dot $Q$) describing single-spin coherence decay. With $\ket{\Psi_{\pm}}$ states as example, using Eq.~(\ref{eq:rhocorr_separable}), and defining $\Delta p \! \doteq \! p_{A}-p_{B} \! = \! \Delta h^{z}/\mathcal{A}J$, we have
\beq
\frac{\rho_{+-,-+}(t)}{\rho_{+-,-+}(0)} =\sum_{p_{A}} w(p_{A}; \Delta p) K^{A,+-}_{a,p_{A}}(t)K^{B,-+}_{a,p_{B}=p_{A}-\Delta p}(t) \,\, , \label{eq:rhoKAKB}
\eeq
where the $K^{Q,\sigma\bar{\sigma}}_{a,p_{Q}}$ are the single-dot dephasing functions, i.e.~the relevant two-spin coherence is an appropriately weighted average over results of calculations assuming \textit{uncorrelated} baths. 
The qualitative properties of this expression are most easily seen in the UC approximation. Using the notation from Eqs.~(\ref{eq:rhocorr}) and (\ref{eq:rhocorr_separable}) we have $M_{Q}(p_{Q}) \! = \! \sum_{j_{Q}\geq |JN_{Q}p_{Q}|} n_{j_{Q}}$ and 
\beq
w(p_{A}; \Delta p) = \frac{M_{A}(p_{A}) M_{B}(p_{B} \! = \! p_{A} \! - \! \Delta p)}{\sum_{p_{A}} M_A(p_{A}) M_{B}(p_{B} \! = \! p_{A} \! - \! \Delta p)} \,\, . \label{eq:wAB}
\eeq
The behavior of the expression (\ref{eq:rhoKAKB}) is then determined by the fact that according to an approximate formula (\ref{eq:nj_approx}) the degeneracy factors $n_{j_{Q}}\! \sim \! j_{Q} \exp(-2j_{Q}^{2}/N)$ decrease very rapidly with $j_{Q}$ for $j_{Q} \! \gg \! \sqrt{N_{Q}/2}$. Consequently, the weights from Eq.~(\ref{eq:wAB}) are maximized when $p_{A} \! = \! \Delta p N_{B}/(N_{A}+N_{B})\!  \doteq \! \bar{p}_{A}$ (to which  $p_{B}\! = \! -\Delta p N_{A}/(N_{A}+N_{B}) \! \doteq \! \bar{p}_{B}$ corresponds). This is illustrated in Fig.~\ref{fig:CorBaths_weights}, where $w(m_{A}; \Delta m)$ with $m_{A} \! = \! p_{A}N_{A}J$ is plotted.
While the narrowed bath states corresponding to these $\bar{p}_{Q}$ have dominant influence on the coherence decay, in Eq.~(\ref{eq:rhoKAKB}) we still sum over a set of states with $p_{Q}$ close to these values. This summation leads to averaging out (on timescale of $\sim \! T_{2}^{*}$) of fast oscillations of the NFID coherence signal discussed previously. 
Consequently, the best approximation for the coherence (and entanglement) decay is given by
\beq
\frac{\rho_{+-,-+}(t)}{\rho_{+-,-+}(0)} \approx  K^{A,+-}_{a, \bar{p}_{A}}(t)K^{B,-+}_{a,\bar{p}_{B}}(t) \,\, , \label{eq:Ccorr_approx}
\eeq
where $K^{Q,\sigma\bar{\sigma}}_{a,\bar{p}_{Q}}$ are the oscillation-free functions from Eq.~(\ref{eq:NFID_RDTp}). Note also that while the simplicity of this derivation relied on assuming that the relevant $m_{Q} \! \gg \! \sqrt{N_{Q}/2}$, it is easy to check that for $\Delta p \! =\! 0$ Eq.~(\ref{eq:rhoKAKB}) is dominated by $m_{Q} \! \lesssim \! \sqrt{N_{Q}}$, so that we can approximate the decay function by a product of two zero-polarization single-dot $K^{\sigma\bar{\sigma}}_{a}$ functions from Eq.~(\ref{eq:NFID_RDT}).

\begin{figure}[t]
\includegraphics[width=\linewidth]{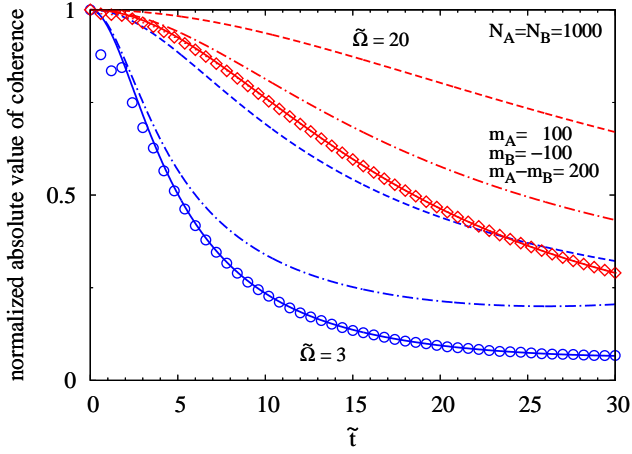}
\caption{(Color online)~Normalized absolute values of single-qubit and two-qubit coherences (for initial $\ket{\Psi_{\pm}}$ state) calculated for two dots with $N_{A}\! =\! N_{B} \! =1000$ in fields $\tilde{\Omega} = 3$ (in blue) and $\tilde{\Omega} = 20$ (in red). Dashed and dot-dashed lines correspond to coherences of qubits A and B, respectively, calculated using Eq.~(\ref{eq:NFID_RDTp}) assuming $p_{A}\! = \! 0.2$ and $p_{B} \! = \!-0.2$. Since $h^z_{Q}$ is enhancing (suppresing) the total qubit splitting for dot A (B), the decays of these coherences are visibly distinct, especially for lower value of external field. 
Solid lines correspond to the absolute value of  $K_{+-,-+}(t)\! = \! K_a^{A,+-}(t) \: K_a^{B,-+}(t)$  function from Eq.~(\ref{eq:Ccorr_approx}), i.e.~the two-qubit coherence $\rho_{+-,-+}(t)$ calculated assuming uncorrelated narrowed baths. Symbols correspond to the absolute value of $\rho_{+-,-+}(t)$ calculated with the UC approach for correlated baths narrowed to $\Delta p \! =\! 0.4$. The agreement of the latter with the solid lines is very good for $\tilde{t}\! \gtrsim \! 1$ (at shorter times the UC solution exhibits oscillations, see solid lines in Fig.~\ref{fig:C_narrowed_long}).
}
\label{fig:NSvsCorBaths_p02}
\end{figure}

The above derivation explains the similarity of the entanglement decay between Fig.~\ref{fig:C_narrowed} and Fig.~\ref{fig:C_correlated}, illustrated in Fig.~\ref{fig:C_narrowed_long}, in which the full UC calculation for correlated baths (with $\Delta h^{z} \! =\! 0$) is compared with an UC calculation assuming uncorrelated baths, each narrowed to $h^{z}_{Q}\! =\! 0$.
Figure \ref{fig:NSvsCorBaths_p02} contains further examples of the accuracy of Eq.~(\ref{eq:Ccorr_approx}) in the case of correlated state with non-zero $\Delta h^{z}$.

\section{Entanglement echo}  \label{sec:echo}
Spin echo protocol leads to a very efficient recovery of single-qubit coherence\cite{Petta_Science05,Koppens_PRL08,Bluhm_NP10} because of quasistatic nature of dephasing caused by a thermal nuclear bath. For electron spins in gated QDs, it is now well established that at low $B$ fields (when the time scale of $t \! < \! N/\MA$ is of interest) the dynamics of the echo signal is caused by nuclear Larmor precession, and the presence of distinct nuclear species with $\omega_{\alpha}\! \neq \omega_{\beta}$ is crucial
\cite{Cywinski_PRL09,Cywinski_PRB09,Cywinski_PRB10,Bluhm_NP10,Neder_PRB11} for times larger than $1/\omega_{\alpha\beta}$ (where $\omega_{\alpha\beta} \! \doteq \! \omega_{\alpha}-\omega_{\beta}$).

The same degree of efficiency is expected when we apply the echo procedure to two entangled qubits, especially when $\Omega \! \gg \! \so$, so that the influence of the nuclear baths amounts mostly to pure dephasing. For a single spin, the echo is refocusing the coherence of superpositions of $\hat{S}^{z}$ with $\pi$ pulse about $x$ or $y$ axis -- both these pulses lead to exchange of amplitudes between $\ket{+}$ and $\ket{-}$ states.\footnote{To be precise, the $\pi_{x}$ and $\pi_{y}$ pulses correspond to $-i\hat{\sigma}_{x}$ and $-i\hat{\sigma}_{y}$ operators, the application of which introduces additional phase factors. These can be easily taken into account when relating the echoed state to the original one.}  The recovery of coherence of superpositions of two-qubit states can be done in two distinct ways. When dealing with $\ket{\Psi_{\pm}}$ Bell states, we can recover the $\rho_{+-,-+}$ coherence by using the two-qubit SWAP gate, and this method was in fact used to protect the $\ket{\Psi_{-}}$  state 
in a DQD.\cite{Petta_Science05,Bluhm_NP10} This method is inefficient for $\ket{\Phi_{\pm}}$ states, which are invariant under the SWAP operation. One can however try to recover the coherence of \textit{any} of the Bell states by simultaneous application of $\pi_{x/y}$ pulses to the two qubits.\cite{Shulman_Science12,Dolde_NP13,LoFranco_PRB14} This operation works in complete analogy to the single qubit echo: for every Bell state, the $\pi$ pulses are exchanging the amplitudes of the two relevant states.

\begin{figure}
\includegraphics[width=\linewidth]{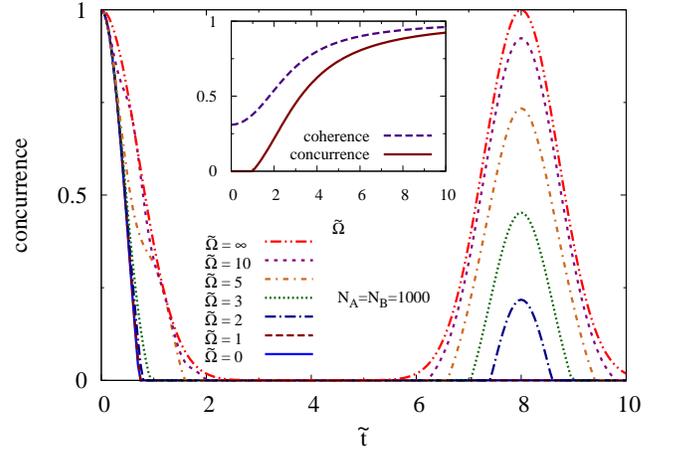}
\caption{(Color online)~Concurrence as a function of time in the presence of echo $\pi$ pulse at $\tilde{t}\! = \! 4$ for different values of magnetic field, calculated for an initial Bell state. The calculation is performed within the UC model assuming a single nuclear species, on timescale of $t \! \ll \! \omega_{\alpha}$ (so that the values of $\omega_{\alpha}$ are irrelevant and in the calculation we put them equal to zero). Note that for $\tilde{\Omega} \! = \! 0$ and $1$ the entanglement does not revive at the echo time of $\tilde{t} \! =\! 8$. For larger $\tilde{\Omega}$ the entanglement is indeed revived by the echo procedure and its peak value grows with increasing $\tilde{\Omega}$. Inset: absolute value of normalized two-qubit coherence vs entanglement at the time of maximum of the echo signal as a function of $\tilde{\Omega}$. At lowest magnetic fields the amount of recovered coherence is not large enough to lead to a recovery of entanglement. 
} \label{fig:echo_real_time}
\end{figure} 

The fact that the simultaneous application of two single-qubit (i.e.~local) operations can lead to recovery of entanglement is quite intuitive when the interaction is approximately of pure dephasing kind and the bath is slow -- such an echo procedure was recently used in experiments on entanglement of spin qubits.\cite{Shulman_Science12,Dolde_NP13} At first sight one could raise an objection that an application of two local operations at time $t \! =\! \tau$, when no entanglement might be present among the two qubits (see Fig.~\ref{fig:echo_real_time}), leads to appearance of large entanglement at later time $t \approx 2\tau$, apparently violating the paradigm that local operations and classical communication (LOCC) cannot increase the amount of entanglement.\cite{Plenio_QIC07,Horodecki_RMP09,Aolita_RPP15} However, one has to remember that the LOCC paradigm rests on assumption that the two-qubit evolution is described by a completely positive (CP) map. For two qubits interacting with a quasistatic pure 
dephasing bath, 
only the evolution starting from the point in time when the qubits are uncorrelated with the baths, is in fact CP. The evolution from $t \! = \! \tau$ onward is \textit{not} CP, because the states of the qubits and the environments at this time are correlated 
(see Ref.~\onlinecite{Buscemi_PRL14} and references therein, and Ref.~\onlinecite{DArrigo_AP14} for a discussion focused on the case of echo), and this correlation is in fact the reason for diminished entanglement between the qubits at this moment. Taking the time of initialization of two-qubit state as a reference point, the echo procedure does not result in any increase of entanglement.

\begin{figure}
\includegraphics[width=\linewidth]{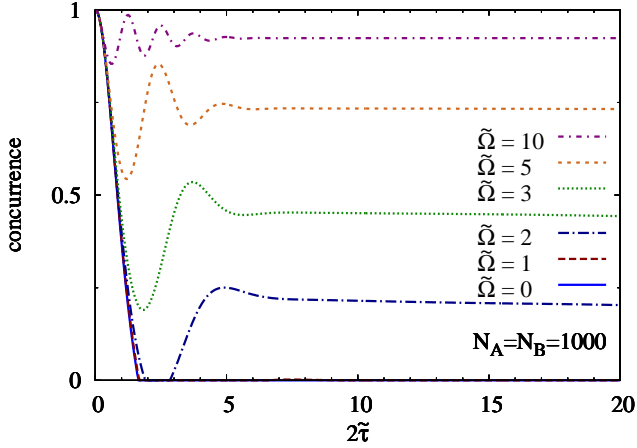}
\caption{(Color online)~Concurrence at the maximum of echo-induced revival as a function of total echo sequence time for various magnetic fields calculated using the UC approach.
The electron spins are initially in one of the Bell states and interact with two separate baths consisting of a single nuclear species.
} \label{fig:echo_homo}
\end{figure} 

\begin{figure}
\includegraphics[width=\linewidth]{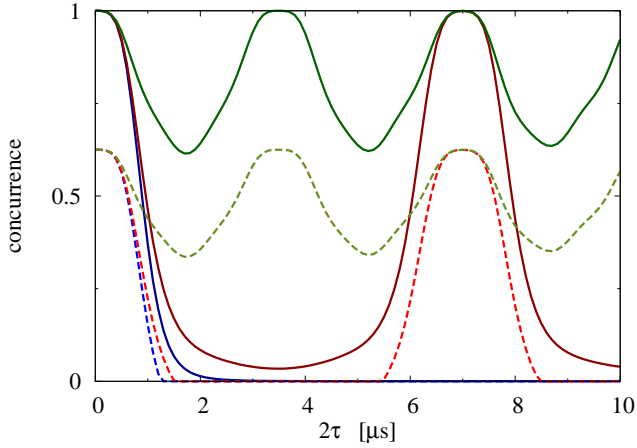}
\caption{(Color online)~Concurrence at the maximum of echo-induced revival as a function of total echo sequence time for GaAs QDs (with $N_{A}\! = \! N_{B}\! = \! 10^{6}$) at $B$ field of $50$, $100$ and $200$ mT (the amplitude of the oscillations and their apparent period grows with decreasing $B$). Solid lines are results for Bell states, while the dashed lines are results for a Werner state with $p\! = \! 3/4$.
} \label{fig:echo_RDT}
\end{figure} 

We focus then on the two-qubit echo performed with synchronized single-qubit $\pi$ rotations. First let us look at the regime of lowest magnetic fields, $\Omega_{Q} \! \lesssim \! \so$, and focus on time scale of $t \! \ll \! 1/\omega_{\alpha\beta}$ on which we can disregard the existence of distinct $\omega_{\alpha}$ splittings of various nuclear species. We can then use the single-species UC approach to spin echo signal from Secs.~\ref{sec:evo_SE} and \ref{sec:UC}. In Fig.~\ref{fig:echo_real_time} we show the time-dependence of concurrence in a ``real-time'' version of echo protocol: for $t \! < \tau_{1}=\! 4T_{2}^{*}$ we show the free evolution decay of entanglement, and for later times we show the evolution of $C(t\!=\! \tau_{1}+\tau_{2})$ after application of $\pi$ pulses at $\tau_{1}$.  For large enough $\tilde{\Omega}$ at $\tau_{2} \! \approx \! \tau_{1}$ we obtain a partial entanglement recovery. While the two-qubit coherence is always partially recovered at the echo time 
(see the inset of Fig.~\ref{fig:echo_real_time}), the non-zero entanglement reappears only above $\tilde{\Omega} \! \approx \! 1$. This should not be surprising: for $\tilde{\Omega} \! < \! 1$ the diagonal elements of the two-qubit density matrix are strongly perturbed, and partial recovery of coherence might not be enough for entanglement revival to occur (note that according to Eq.~(\ref{eq:CX}) for such a revival to happen the echoed coherence needs to fulfill $|\rho_{ab}| \! > \! \sqrt{\rho_{cc}\rho_{dd}}$, where $\rho_{cc}$ and $\rho_{dd}$ are the occupations created by bath-induced qubit dynamics).

In Fig.~\ref{fig:echo_homo} we present a more customary plot of dependence of peak echo amplitude on the total sequence time $2\tilde{\tau}$. In an effectively homonuclear system at higher fields the entanglement exhibits initial partial visibility loss of amplitude $\propto \! 1/\tilde{\Omega}^{2}$ for $\tilde{\Omega} \! \gg \! 1$, and then it stays constant. 
If the nuclear bath was really homonuclear, then at longer times a small-amplitude oscillation of $C(t)$ with frequency $\omega$ would appear, as in the single-spin echo case.\cite{Cywinski_PRB10} Then the entanglement would decay towards zero at much longer times due to dipolar-induced bath dynamics causing dephasing, as described in Sec.~\ref{sec:beneath}. This is the well-known effect of echo sequence removing almost all the influence of transverse Overhauser fields on qubits' coherence when dealing with a homonuclear bath.\cite{Yao_PRB06,Cywinski_PRB10}

In a III-V QDs the presence of multiple nuclear species completely changes\cite{Cywinski_PRL09,Cywinski_PRB09,Bluhm_NP10,Neder_PRB11} the echo signal for $t \! > \! 1/\omega_{\alpha\beta}$. 
Then, assuming only $t\! \ll \! N/\MA$, we can use the RDT (see Sec.~\ref{sec:RDT}) to calculate the coherence dynamics due to presence of multiple nuclear species (each uniformly coupled to the electron). We take the single-qubit $K^{\sigma\bar{\sigma}}_{a}(t)$ function from Refs.~\onlinecite{Cywinski_PRL09,Cywinski_PRB09,Neder_PRB11}:
\beq
K^{\sigma\bar{\sigma}}_{a}(t) = \left( 1 + \sum_{\alpha > \beta} a_{\alpha}a_{\beta}n_{\alpha}n_{\beta}\frac{4\MA^{2}_{\alpha}\MA^{2}_{\beta}}{N^{2}\Omega^{2}\omega^{2}_{\alpha\beta}} \sin^{4}\frac{\omega_{\alpha\beta}t}{4} \right)^{-1} \,\, ,
\eeq
where $a_{\alpha} \! \doteq \! \frac{2}{3}J_{\alpha}(J_{\alpha}+1)$. Using these functions for realistic parameters of GaAs quantum dots with $N_{A}\! = N_{B} \! = 10^{6}$ we obtain results for concurrence dynamics of Bell and Werner states shown in Fig.~\ref{fig:echo_RDT}. Since the RDT is based on pure dephasing approximation, for an initial Bell state there is no ESD, and apparent vanishing of entanglement (see the line corresponding to the lowest $B$ field) is simply related to $C(t)$ becoming very small (but strictly speaking non-zero) for some periods of time. On the other hand, when a Werner state is initialized, at low enough $B$ fields we see events of entanglement death and revival (and also a diminished maximal value of entanglement at times of nearly-perfect returns of the signal).

\section{Projection on a singlet and average teleportation fidelity as witnesses of decaying entanglement} \label{sec:witnesses}
We finally revisit the other methods of verifying the existence, and quantifying the amount of entanglement, which were introduced in Sec.~\ref{sec:quantification}: the measurement of an entanglement witness related to projection on one of the Bell states (say, singlet $\ket{\Psi_{-}}$) and the measurement of average quantum teleportation fidelity.

\begin{figure}
\includegraphics[width=\linewidth]{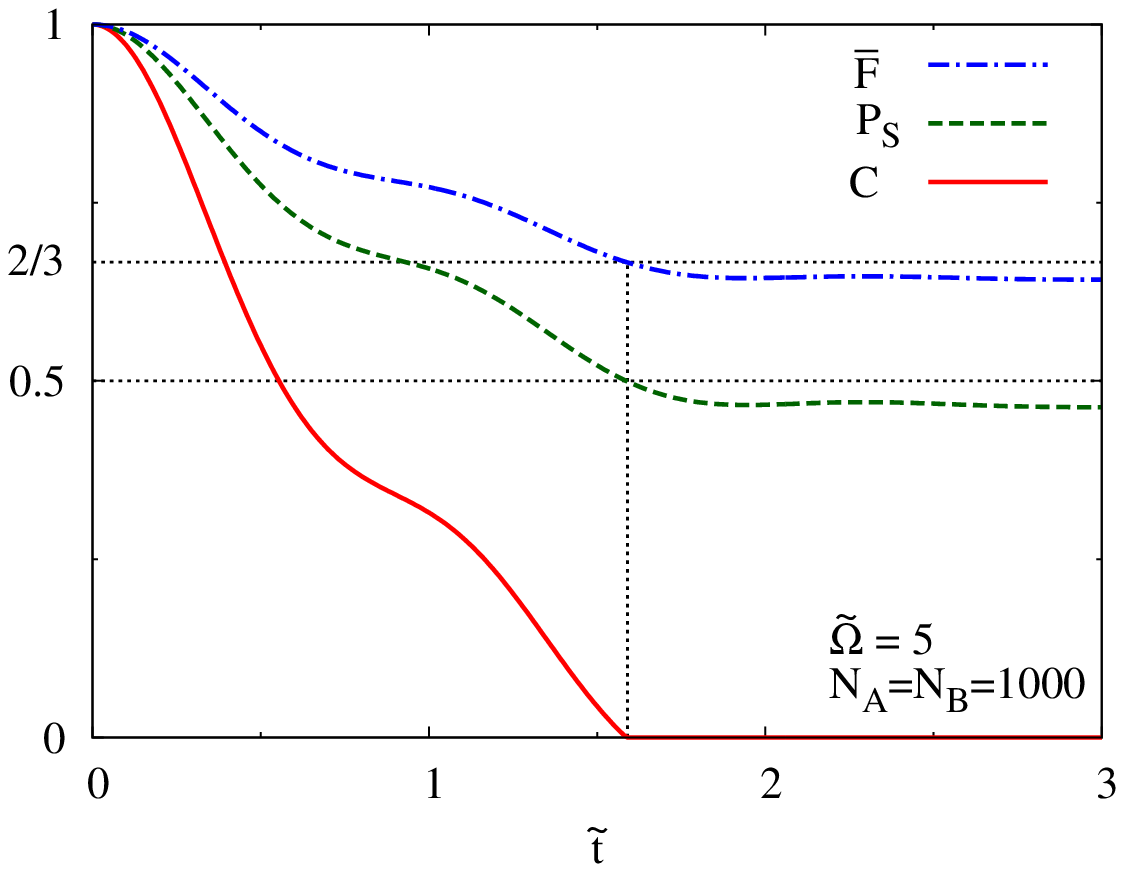}
\caption{(Color online)~Concurrence $C(\tilde t)$, projection on singlet $P_S(\tilde t)$, and average fidelity of teleportation $\bar{F}(\tilde t)$
for two electron spins initially being in a singlet state $ | \Psi_{-} \rangle$ for the case of interaction with two separate nuclear spin baths in high-temperature states calculated using the UC approach at $\tilde{\Omega} \! =\! 5$. The vertical dotted line marks the time at which the state becomes disentangled, while the horizontal dotted lines at $0.5$ and $2/3$ correspond to values at which $P_{S}(\tilde t)$ and $\bar{F}(\tilde t)$, respectively, cease to indicate the presence of entanglement.
Time is in units of two-spin $T_{2}^{*}$ defined in Eq.~(\ref{eq:T2star2}), and the dimensionless Zeeman splitting is $\tilde{\Omega} \! \doteq \! \Omega T_{2}^{*}$.
}
\label{fig:concurrence_projection_fidelity}
\includegraphics[width=\linewidth]{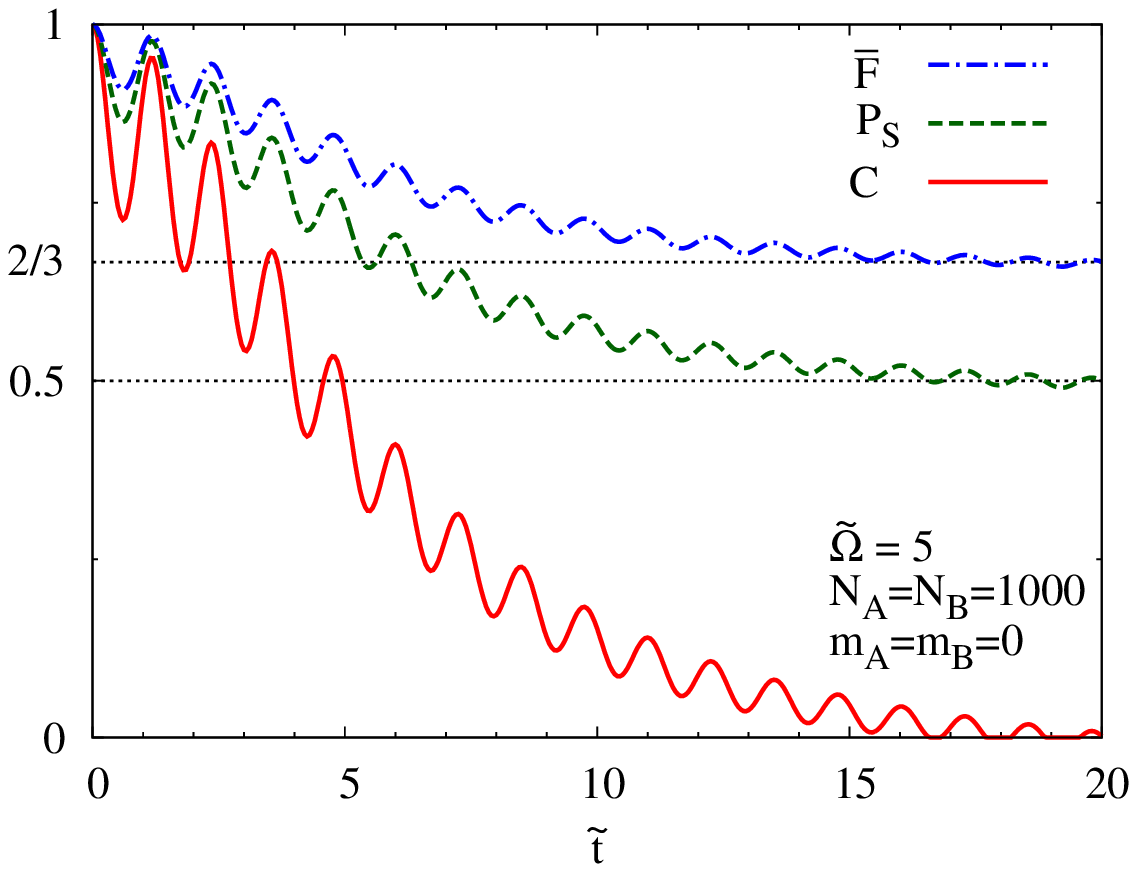}
\caption{(Color online)~The same as above, only for narrowed state of $A$ and $B$ baths (each having $h^{z}_{Q}\! = \! 0$).
} \label{fig:concurrence_projection_fidelity_NS}
\end{figure} 

The effectiveness of both of these methods relies on having some prior knowledge of the character of the mixed entangled state $\hat{\rho}(t)$ at the time of employing them. Let us then summarize what we have learned about the general form of $\hat{\rho}(t)$, assuming that one of Bell states was initialized. The $X$ form of the state is preserved and furthermore only the initially non-zero coherence remains finite when free evolution is considered (in the case of spin echo this statement is only approximately true, since the second coherence is in fact generated during the evolution, but its magnitude is suppressed as $1/\tilde\Omega^{2}$ at large fields). Initially zero occupations become finite, of magnitude $\propto \! 1/\tilde\Omega^{2}$, and their values are approximately equal. The initially non-zero coherence is of course decaying due to the dominantly pure-dephasing influence of the bath. As a result, the state $\hat{\rho}(t)$ is a Werner-like state with diminished coherence,
\beq
\hat{\rho}(t) \approx \begin{pmatrix}
\frac{1-p(t)}{4} & 0 & 0 & 0 
\\
0 & \frac{1+p(t)}{4} & -|\rho_{23}(t)|\text{e}^{-i\gamma(t)} & 0
\\
0 & -|\rho_{23}(t)|\text{e}^{i\gamma(t)} & \frac{1+p(t)}{4} & 0
\\
0 & 0 & 0 & \frac{1-p(t)}{4}
\end{pmatrix}  \label{eq:WL}
\eeq  
where we assumed that the initial state was $\ket{\Psi_{-}}$, $p(t)\! \approx \! 1-2\big(K^{A}_{b}(t)+K^{B}_{b}(t)\big)$ (we focus now on free evolution and assume $K^{Q,+}_b \! \approx \! K^{Q,-}_b$), and $\gamma(t)$ is a possibly non-trivial phase. The latter feature has not received any attention until now, because the concurrence of the above state depends only on the modulus of $\rho_{23}(t)$. In the above example, for a thermal bath state and for the case of free evolution we have $\gamma(t) \! = \! (\Omega_{A}-\Omega_{B})t$, so that a non-trivial phase appears when the two dots are in a magnetic field gradient. In the case of narrowed baths, apart from the fact that $h^z_{Q}$ contribute now to $\Omega_{Q}$, we have a more non-trivial effect. Even for $h^{z}_{Q}\! =\! 0$ and no magnetic field gradient, due to the presence of the phase term in Eq.~(\ref{eq:NFID_RDT}) we have $\gamma(t) \! =\! \arctan(t/\tau_{A}) - \arctan(t/\tau_{B})$, where $\tau_{Q}$ given in Eq.~(\ref{eq:tauNFID}) depends on the bath 
size $N_{Q}$, leading to non-trivial phase dynamics in the case of non-identical QDs. Analogous considerations for initial $\ket{\Phi_{\pm}}$ state give us $\gamma(t) \! = \! \Omega_{A}+\Omega_{B}$ in the thermal bath case, so that a fast phase dynamics is always present for these states at finite magnetic field. Note that $\gamma\! =\! 0$ when two-spin echo protocol is employed. 

For $\gamma \! =\! 0$, the witness proposed in Sec.~\ref{sec:quantification}, $\hat{w}_{S} \! =\! 1 - \hat{P}_{S}$, is in fact an optimal one,\cite{Guhne_PRA02} since it always detects entanglement when it is present
(see also discussion of this fact specific to the maximally mixed bath case in Ref.~\onlinecite{Mazurek_PRA14}).

In fact it is trivial to check that $\mean{\hat{w}_{S}} \! =\! -\frac{1}{2}C(t)$. In the following we will focus on the expectation value  $P_{S}(t) = \text{Tr} \big( \hat \rho(t)\hat{P}_{S} \big) $ remembering that $P_{S}(t) \! > \! 1/2$ signifies entanglement. However, for $\gamma(t) \! \neq \! 0$, $P_{S}(t) \! < \! 1/2$ does not mean that the state is necessarily separable. We have $P_{S}(t) \! = \! |\rho_{23}(t)|\cos \gamma(t) + (p(t)+1)/4$, while $C(t)\! = \! \text{max}[2|\rho_{23}(t)|-(1-p(t))/2,0]$.  

A similar situation arises when using average quantum teleportation fidelity as an entanglement witness. With the state initialized as $\ket{\Psi_{-}}$ we perform the teleportation protocol designed to work perfectly when $\hat{\rho}$ corresponds to a pure state $\ket{\Psi_{-}}$, and assuming only that $\hat{\rho}(t)$ is of the $X$ form we obtain the fidelity of teleportation of qubit $C$ in state $\alpha |+\rangle + \beta |-\rangle$:
\begin{align}
F_{\alpha,\beta} &= 2 |\alpha|^2 |\beta|^2 (\rho_{11} - \rho_{22} - \rho_{33} + \rho_{44})
     - 4 |\alpha|^2 |\beta|^2 \text{Re} \rho_{23} \nonumber \\
    &+ \rho_{22} +\rho_{33}
     - 4 \text{Re} \Big(\alpha^2 (\beta^*)^2\Big) \text{Re} \rho_{14}.
\end{align}
Using now the $A$ and $B$ qubits described by density matrix $\hat{\rho}(t)$ from Eq.~(\ref{eq:WL}) we obtain
\begin{align}
F_{\alpha,\beta}(t) &= -2p(t) |\alpha\beta|^2 + 4 |\alpha\beta|^2 |\rho_{23}(t)|\cos\gamma(t) +\frac{1+p(t)}{2} \,\, .\label{eq:F2} 
\end{align}
Note that the fidelity teleportation of basis states, $|+\rangle$ and $|-\rangle$, does not depend on the presence of coherence, and it is simply equal to $(1+p)/2$. When the initialized state is a Werner state, measuring the fidelity of teleportation of one of the basis states immediately after the initialization gives us the value of $p$. 

\begin{figure}
\includegraphics[width=\linewidth]{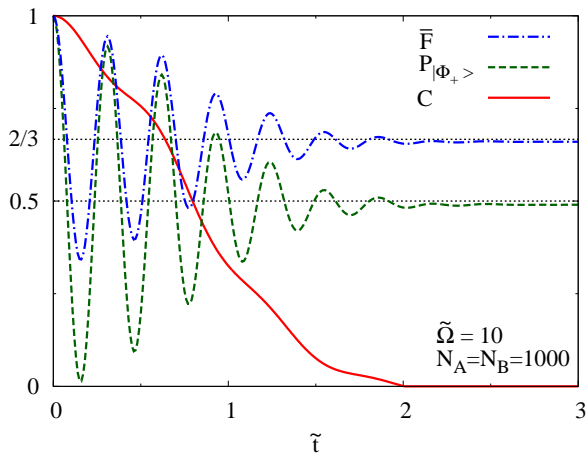}
\caption{(Color online)~Concurrence $C(\tilde t)$, projection on $\ket{\Phi_{+}}$ Bell state $P_{\ket{\Phi_{+}}}(\tilde t)$, and average fidelity of teleportation $\bar{F}(\tilde t)$ for two electron spins initially being in $ | \Phi_{+} \rangle$ state for the case of interaction with two separate nuclear spin baths in thermal states calculated using the UC approach for  $\tilde{\Omega} \! =\! 10$.
} \label{fig:concurrence_projection_fidelity_Phi}
\end{figure} 

The fidelity averaged over the teleported states is
\beq
\bar{F}(t) = \frac{2}{3}|\rho_{23}(t)|\cos\gamma(t) +\frac{1}{2} + \frac{p(t)}{6} \,\, .
\eeq
When $\gamma\! =\! 0$, we have $\bar{F}(t) = C(t)/3+2/3$, so that when $\bar{F} \! =\! 2/3$ the $\hat{\rho}(t)$ state becomes separable. For non-zero $\gamma(t)$ the average fidelity oscillates between $1/3$ and $1$. Note that $\bar{F} \! < \! 1/2$ means that the teleported state is anticorrelated with the desired state, which is a clear sign of using the wrong assumption about the $\hat{\rho}(t)$ state (i.e.~~using a wrong teleportation protocol).

The time-dependence of both $P_{S}(\tilde t)$ and $\bar{F}(\tilde t)$ in the case of $\gamma\! = \! 0$ is shown for thermal and narrowed bath states in Figs.~\ref{fig:concurrence_projection_fidelity} and \ref{fig:concurrence_projection_fidelity_NS}, respectively. It is clear that these quantities contain the same information as the concurrence. Note that the result for $P_{S}(\tilde t)$ from Fig.~\ref{fig:concurrence_projection_fidelity} corresponds to moderate magnetic field ($\tilde{\Omega}\! = \! 5$), so that it lies between the limits of $\tilde{\Omega} \! \ll \! 1$ and $\tilde{\Omega} \! \gg \! 1$ investigated experimentally and theoretically 
in Ref.~\onlinecite{Petta_Science05}.

An example of the effects of $\gamma(t) \! \neq \! 0$ is shown in Fig.~\ref{fig:concurrence_projection_fidelity_Phi}, in which the entanglement decay of $\ket{\Phi_{+}}$ state is shown. The calculations of $P_{\ket{\Phi_{+}}}(t)$ and $\bar{F}(t)$ obtained with the teleportation protocol assuming that $\ket{\Phi_{+}}$ state is the entangled resource, are analogous to the ones given above for $\ket{\Psi_{-}}$ state.
Both of these quatities are exhibiting oscillations, and only after fitting an envelope to these signals one can obtain the amount of entanglement at given time. Of course, when $\gamma(t)$ is known, one can either employ an experimental procedure which results in projection on (or teleportation with) a ``correct'' state, but with this knowledge one can more easily obtain the smooth decay curves by post-processing of the data.

\section{Conclusion}
We have presented calculations of nuclear-bath-induced entanglement dynamics of two uncoupled quantum dot spin qubits initialized in a Bell-diagonal state. We considered various experimentally relevant states of the bath (including a correlated narrowed state corresponding to a well-defined difference of $h^{z}$ components of Overhauser fields in the two dots) for the cases of free evolution of the qubits and of the two-qubit spin echo, focusing on low and moderately-low magnetic fields, for which the bath can affect the $z$ component of the electron spin.

The main results of the analysis presented in this paper are the following. 
The influence of the bath (especially at magnetic field $B$ larger than the typical Overhauser field) can be approximately described as leading to the two-qubit state becoming Werner-like, i.e.~becoming a mixture of an initial Bell state subjected to pure dephasing, and a completely mixed density matrix with weight $\propto 1/B^2$. The latter dynamically generated admixture leads to appearance of entanglement sudden death at finite time (and possible revival at a later time) when initial Bell state is considered (as it was obtained in earlier work focused on thermal state of the nuclei, Ref.~\onlinecite{Mazurek_PRA14}). The quantum channel describing the hf-induced disentanglement can be thus roughly described as non-Markovian phase damping channel with a small admixture of generalized amplitude damping \cite{Aolita_RPP15} corresponding to infinite-temperature environment. 
For a correlated narrowed state of two baths, we have shown that due to the fact that the number of bath states corresponding to a given nuclear polarization $p$ rapidly diminished with increasing $p$, the two-spin dynamics can be calculated assuming an effectively uncorrelated state of the two baths, with specific polarizations assigned to each of them.

Entanglement can be recovered using a two-spin echo procedure,\cite{Shulman_Science12,Dolde_NP13} in which local operations ($\pi$ pulses) applied to two qubits lead to refocusing of the two-qubit coherence and revival of entanglement. This procedure fails at lowest magnetic fields (smaller and comparable to the typical Overhauser field), at which the refocusing of coherence cannot turn the Werner-like separable state into an entangled one. 

Finally, we have shown that hyperfine-induced entanglement decay of Bell states can be detected and quantified without performing the two-qubit tomography. The entanglement (measured by concurrence) can be {\it faithfully} reconstructed from a measurement of a simple entanglement witness related to projection on the initial entangled state (see also Ref.~\onlinecite{Mazurek_PRA14}) and from the measurement of average quantum teleportation fidelity -- for the Werner-like state generated from the initial Bell state by interaction with the nuclear bath, these two measurements detect the presence of entanglement if and only if it is indeed present.

\section*{Acknowledgements}
We would like to thank R.~Hanson for an enlightening discussion concerning quantum teleportation.
This work is supported by funds of Polish National Science Centre (NCN), grant no.~DEC-2012/07/B/ST3/03616.

\appendix

\section{Analytical solutions in the uniform coupling approximation} \label{app:UC}
In the UC model the dynamics occurs in uncoupled two-dimensional subspaces of $\ket{\sigma,j,m}$ and $\ket{\bar{\sigma},j,m+\sigma}$ states (where $\sigma \! = \! \pm 1$ is the $\hat{\sigma}_{z}$ eigenvalue for the electron spin):
\beq
\text{e}^{-i\hat{H}t} \ket{\sigma,j,m} \equiv a_{jm\sigma}(t) \ket{\sigma,j,m} + b_{jm\sigma}(t) \ket{\bar{\sigma},j,m+\sigma} \,\, .
\eeq
We define the quantities\cite{Barnes_PRB11}
\begin{align}
E_{m\sigma} & \doteq \frac{1}{2}\big((2m+\sigma)\omega - \mathcal{A}/2N \big) \,\, ,\label{eq:E} \\
x_{jm\sigma} & \doteq \mathcal{A}\sqrt{j(j+1)-m(m+\sigma)}/N \,\, ,\label{eq:X} \\
z_{m\sigma} & \doteq \sigma \big( \Omega-\omega+\mathcal{A}(m+\sigma/2)/N \big) \,\, ,\label{eq:Z} \\
v_{jm\sigma} & \doteq \sqrt{x^{2}_{jm\sigma}+z^{2}_{m\sigma}} \,\, ,\label{eq:N} 
\end{align}
using which we have
\begin{align}
a_{jm\sigma}(t) & \doteq \text{e}^{-iE_{m\sigma}t}\left( \cos v_{jm\sigma}t/2 - i \frac{z_{m\sigma}}{v_{jm\sigma}}\sin v_{jm\sigma}t/2 \right ) \,\, , \label{eq:a} \\
b_{jm\sigma}(t) & \doteq -i\text{e}^{-iE_{m\sigma}t}\frac{x_{jm\sigma}}{v_{jm\sigma}}\sin v_{jm\sigma}t/2  \,\, . \label{eq:b} 
\end{align}
The evolution of an electron spin coupled to the nuclei is then described by the functions $K^{\sigma\sigma'}_{a}(t)$ and $K^{\sigma}_{b}(t)$ defined in Sec.~\ref{sec:evolution}:
\begin{align}
K^{\sigma\sigma'}_{a}(t) & = \frac{1}{Z}\sum_{jm} n_{j} p_{m} a_{jm\sigma}(t)a^{*}_{jm\sigma'}(t) \,\, , \label{eq:KaUC} \\
K^{\sigma}_{b}(t) & = \frac{1}{Z}\sum_{jm} n_{j} p_{m} |b_{jm\bar{\sigma}}(t)|^{2} \,\, , \label{eq:KbUC}
\end{align}
where $Z$ is the normalization factor, $p_{m}$ are the appropriate weights, and $n_{j}$ are the degeneracies of subspaces with given $j$. For a high-temperature bath of nuclear spins $J$ we have $Z \! =\! (2J+1)^{N}$ and $p_{m} \! = \! 1$, while for fully narrowed nuclear state with $h^{z} \!= \! m_{0}\mathcal{A}/N$ we have $p_{m} \! = \! \delta_{mm_{0}}$ and $Z\! = \! \sum_{j\geq |m_{0}|} n_{j}$.

The degeneracy factor $n_{j}$ is given for nuclear bath consisting of $N$ spins $\frac{1}{2}$ by\cite{Melikidze_PRB04}
\begin{align}
n_{j} & = \frac{N!}{(N/2-j)!)(N/2+j)!}\frac{2j+1}{N/2+j+1} \label{eq:nj_exact} \\ 
& \approx 2^{N} \frac{4(2j+1)}{\sqrt{2\pi}N^{3/2}} \text{e}^{-2j^2/N} \,\, , \label{eq:nj_approx}
\end{align}
where in the second expression we have assumed $N\! \gg \! 1$ and $j\! \ll \! N/2$. 

When the bath is in a high-temperature state, for $\Omega \! \gg \! \so \sim \mathcal{A}/\sqrt{N}$ we have
\beq
\frac{z_{jm\sigma}}{v_{jm\sigma}} \approx \sigma \left (1-\frac{\so^{2}}{2\Omega_{m\sigma}^{2}} \right ) \,\, , \label{eq:ZN}
\eeq 
where $\Omega_{m\sigma} \doteq \Omega-\omega +\mathcal{A}(m+\sigma/2)/N \! \approx \! \Omega$. The last approximation follows from observation that according to Eq.~(\ref{eq:nj_approx}) the sum in Eqs.~(\ref{eq:KaUC}-\ref{eq:KbUC}) is dominated by terms with $j \lesssim \sqrt{N}$, which also limits the relevant values of $m$. We also approximate
\beq
v_{jm\sigma} \approx \Omega_{m\sigma} + \frac{\mathcal{A}^2}{2N\Omega_{m\sigma}} \,\, , \label{eq:Napp}
\eeq
where the second term can be dropped on timescale $t\! \ll \! \Omega/\so^2$. As a result, we obtain
\beq
K^{\sigma\bar{\sigma}}_{a}(t)  \approx \frac{\text{e}^{-i\sigma\Omega t}}{2^{N}}\sum_{jm} n_{j} \text{e}^{-i\sigma\mathcal{A}mt/N} \approx \text{e}^{-i\sigma\Omega t} \text{e}^{-(t/T_{2,Q}^{*})^{2}}  \,\, , \label{eq:KaUChighT} 
\eeq
where we used the fact that the sum over $j$ and $m$ values can be approximated \cite{Merkulov_PRB02} by an integral $\int P(m) \text{d}m$ with $P(m) \propto \exp(-m^2/2\so^2)$ where $\so$ is given in Eq.~(\ref{eq:sigma}), and the resulting single-dot $T_{2,Q}^{*}$ time is given by $\sqrt{2}/\so$. Note that a partial narrowing of the nuclear distribution (i.e.~removal of certain ranges of $m$ values from the sums), resulting in a diminished value of $\so$, leads to the above formula for  $K^{\sigma\bar{\sigma}}_{a}(t)$, only with increased $T_{2,Q}^{*}$, as long as the timescale of interest $t\!  \lesssim \! T_{2,Q}^{*} \! \ll \! \Omega/\so^{2}$. 

In order to obtain an approximation for $K^{\sigma}_{b}(t)$ in this case we have to make an approximation (to the leading order in $\sigma^2/\Omega^{2}$),
\beq
\frac{x^{2}_{jm\sigma}}{v^{2}_{jm\sigma}} \approx \frac{\mathcal{A}^2}{N^{2}\Omega^2} \big( j(j+1)-m(m+\sigma) \big) \,\, ,
\eeq
in Eq.~(\ref{eq:b}), giving
\beq
K^{\sigma}_{b}(t) \approx \frac{1}{2^{N}}\sum_{j,m}n_{j} \frac{\mathcal{A}^2}{N^{2}\Omega^2} \big( j(j+1)-m(m+\sigma) \big) \sin^{2}v_{jm\sigma}t/2 \,\, . \label{eq:KbUCfull}
\eeq
Using $v_{jm\sigma} \! \approx \! \Omega_{m\sigma}$ (valid at $t \! \ll \! T_{2}^{*} \Omega/\sigma$) we see that the oscillatory terms in the above equation average out to $1/2$ for $t\! \gg \! T_{2}^{*}$. At these long times, using the relation \cite{Barnes_PRB11}
\beq
\frac{1}{2^{N}}\sum_{j,m}n_{j} \big( j(j+1) -m(m+\sigma) \big) = N/2 \,\, ,
\eeq
we obtain
\beq
K^{\sigma}_{b}(t\! \gg \! T_{2,Q}^{*}) \approx \frac{\mathcal{A}^2}{4N\Omega^2} = \frac{2}{\tilde{\Omega}^2_{Q}} \,\, , \label{eq:KbUChighT}
\eeq
where $\tilde{\Omega}_{Q} \! \doteq \! \Omega T^{*}_{2,Q}$.

Let us perform an analogous approximate analytical analysis in the case of fully narrowed nuclear bath. Defining $\Omega_{m} \! \doteq \! \Omega-\omega+ \mathcal{A}m/N$ we approximate now 
\begin{align}
z_{m\sigma} & \approx  \sigma\Omega_{m} \,\, ,\\
\frac{z_{m\sigma}}{v_{jm\sigma}} & \approx \text{sgn}(\Omega_{m})\left( 1 - \frac{x^{2}_{jm\sigma}}{2\Omega^{2}_{m}} \right ) \,\, , \\
\frac{x_{jm\sigma}}{v_{jm\sigma}}  & \approx  \frac{x_{jm\sigma}}{|\Omega_{m}|}   \,\, , 
\end{align}
which leads to
\begin{align}
K^{\sigma\bar{\sigma}}_{a}(t) & \approx \frac{\text{e}^{-i\sigma\omega t}}{Z_{m}}\sum_{j\geq |m|} n_{j} \text{e}^{-\frac{i}{2}\text{sgn}(z_{m\sigma})(v_{jm\sigma}+v_{jm\bar{\sigma}})t}  \nonumber\\
& \times \prod_{\sigma = \pm} \! \left( \! 1 + \frac{x^{2}_{jm\sigma}}{4\Omega^{2}_{m}}(\text{e}^{i\text{sgn}(z_{m\sigma})v_{jm\sigma}t}-1)\! \right) \,\, . \label{eq:KaUCNfull}
\end{align}
In the above formula the terms in the second line are responsible for the fast oscillations visible in entanglement decay signal in Fig.~\ref{fig:C_narrowed} (remember that concurrence of a decohered Bell state, when not being very small, is proportional to one of the two-qubit coherences, given by a product of two $K^{\sigma\bar{\sigma}}_{a}(t)$ functions). A quick calculation shows that the amplitude of these oscillations is $\approx 8/\tilde{\Omega}^2$, where $\tilde{\Omega} \! \doteq \! \Omega T_{2}^{*}$, and the two-dot $T_{2}^{*}$ time is defined in Eq.~(\ref{eq:T2star2}).

The first term in Eq.~(\ref{eq:KaUCNfull}) gives the envelope of the decay. Writing $v_{jm\sigma} \! \approx \! |\Omega_{m}| + x^{2}_{jm\sigma}/2|\Omega_{m}|$, we arrive at
\begin{align}
K^{\sigma\bar{\sigma}}_{a}(t) & \approx \text{e}^{-i\sigma(\Omega+\mathcal{A}m/N)t} \frac{1}{Z_{m}} \sum_{j\geq |m|} n_{j} \nonumber\\
&  \exp \! \left( \! -i\sigma\frac{\mathcal{A}^2}{2N^{2}\Omega_{m}} \big(j(j+1)-m^{2} \big) t  \right ) \,\, . \label{eq:KaUCN}
\end{align}
In Ref.~\onlinecite{Barnes_PRB11} it was shown how Eq.~(\ref{eq:NFID_RDT}) follows from the above formula when $m \! = \! 0$. Note that with the above approximation for $v_{jm\sigma}$, the timescale of validity of this result is $t \! \ll \! 4|\Omega_{m}|^{3}/x^{4}_{jm\sigma} \approx 4(N/\mathcal{A})\cdot(\Omega_{m}/\mathcal{A})\cdot\delta_{m}^{-2}$, where $\delta_{m} \! \doteq \! \mathcal{A}/\sqrt{N}\Omega_{m}$, and $\delta_{m} \! \ll \! 1$ is the necessary condition\cite{Cywinski_PRB09,Cywinski_PRB10} for applicability of the effective Hamiltonian approximation leading to Eq.~(\ref{eq:NFID_RDT}). This timescale exceeds $N/\mathcal{A}$ (the time on which the box model is expected to be applicable at all) when $\delta_{m} \! \ll \! (2/N)^{1/6}$, which is not much more restrictive than $\delta_{m} \! \ll \! 1$ for $N \! \geq \! 10^{6}$. 

We can also derive an expression for $K^{\sigma}_{b}(t)$ analogous to the one from Eq.~(\ref{eq:KbUCfull}). Due to the lack of summation over $m$ the oscillations in $K^{\sigma}_{b}(t)$ cancel out at longer times, for $t \! \gg \! \tau_{Q}$ where $\tau_{Q}$ is the characteristic decay timescale of $K^{\sigma\bar{\sigma}}_{a}(t)$ given in Eq.~(\ref{eq:tauNFID}). The value at which $K^{\sigma}_{b}(t)$ saturates is also given by Eq.~(\ref{eq:KbUChighT}), only with $\Omega$ replaced by $\Omega_{m}$.


%

\end{document}